\documentclass[12pt,letterpaper]{article}

\usepackage{appendix}

\usepackage{subcaption}

\usepackage{setspace} 
\usepackage{footmisc}
\usepackage{amsfonts}
\usepackage{amsmath}
\usepackage{theorem}

\usepackage{multirow}

\usepackage[top=2.5cm, bottom=2.5cm, left=2.5cm, right=2.5cm]{geometry}

\usepackage{amsmath}
\usepackage{xcolor}

\newcommand\bovermat[2]{%
	\makebox[0pt][l]{$\smash{\overbrace{\phantom{%
					\begin{matrix}#2\end{matrix}}}^{\text{#1}}}$}#2}

\usepackage{pifont}

\usepackage{url}
\usepackage[T1]{fontenc}

\usepackage{setspace}
\usepackage{footmisc}

\usepackage[natbibapa]{apacite}
\usepackage{amssymb}
\setcitestyle{aysep={}} 

\usepackage{url}
\usepackage{hyperref}

\usepackage{mathtools}
\usepackage{amsmath}
\usepackage[latin2]{inputenc}
\usepackage{graphicx}
\usepackage{subcaption}
\usepackage{multirow}
\usepackage{esvect}
\usepackage{afterpage}

\usepackage{lipsum}
\usepackage{booktabs,array,dcolumn}

\usepackage{adjustbox}
\usepackage{graphicx}
\usepackage{pdflscape}

\usepackage{caption}
\usepackage[labelfont=bf]{caption}
\usepackage[figurename=FIGURE]{caption}
\usepackage[tablename=TABLE]{caption}

\let\ampersand\&
\renewcommand*\&{and}

\makeatletter
\def\@seccntformat#1{\@ifundefined{#1@cntformat}%
	{\csname the#1\endcsname\space}
	{\csname #1@cntformat\endcsname}}
\newcommand\section@cntformat{\thesection.\space}       
\newcommand\subsection@cntformat{\thesubsection.\space} 
\makeatother

\usepackage{atbegshi}
\AtBeginDocument{\AtBeginShipoutNext{\AtBeginShipoutDiscard}}

\begin{document}

\newpage
	\setcounter{page}{1}	
	\begin{titlepage}
		\date{}
		
		\begin{flushleft}
			
			\title{\Large  \textbf{Historical trend in educational homophily: U-shaped or not U-shaped?    
					Or, how to set a criterion to choose a criterion?}}

		\end{flushleft}
		
		\begin{flushleft}
						\singlespacing{\author{Anna NASZODI*} }
			
\thanks{\textit{Email}: anna.naszodi@gmail.com. 
	This paper is being distributed to economists, sociologists and demographers solely to stimulate discussion and elicit comments.   
	\textit{Acknowledgments}: The author acknowledges the U.S. Census Bureau, the source of all the underlying data in IPUMS. Also, she acknowledges the Joint Research Centre (JRC) of the European Commission, as she started to work on this project while being affiliated with the JRC. \textit{Data and code}: Mendeley Data, doi: 10.17632/7cm6673vg8.1.}

		\end{flushleft}

		\maketitle
		\setcounter{page}{1}
		
		\noindent 
		
		\singlespacing{\textbf{Abstract:}  
			
			Measuring changes in overall inequality between different educational groups is often performed by quantifying variations in educational marital homophily across consecutive generations. However, this task becomes challenging when the education level of marriageable individuals is generation-specific. To address this challenge, various indicators have been proposed in the assortative mating literature.
			
			In this paper, we review a set of criteria that indicators must satisfy to be considered as suitable measures of homophily and inequality. Our analytical criteria include robustness to the number of educational categories distinguished and the negative association between intergenerational mobility and homophily. Additionally, we also impose an empirical criterion on the identified qualitative historical trend in homophily between 1960 and 2010 in the US at the national and sub-national levels.

			Our analysis reveals that while  a specific cardinal indicator meets all three criteria, many commonly applied indices do not. We propose the application of this well-performing indicator to quantify the trend in overall		inequality in any country, including European countries, with available population data on couples' education level.}

		\begin{flushleft}
		\end{flushleft}

	\end{titlepage}


\newpage

\section{Introduction}\label{sec:intro}

	Inequality and social cohesion have become a prominent part of the political discourse since the publication of the seminal book by \cite{Piketty2014}.  
	Studying marital homophily by economists and  sociologists, while dating back several decades (see \citealp{Becker1981} and \citealp{Mare1991}),  
	has also attracted renewed attention in the recent years.   
	The reason is that changes in homophily, or, in other words, changes in the degree of marital sorting, are informative about the phenomena of interest:   
    if homophily strengthens, i.e., the degree of sorting increases, then it typically signals an increase in overall inequality between different groups.     
	More precisely, the widening of the social gap between different groups from one generation to the next typically makes 
	people in the later born generation more inclined relative to their peers in the  earlier born generation to choose a partner from their own group.\footnote{Conversely, a change in the opposite direction in aggregate marital/ mating preferences over inter-group versus intra-group romantic relationships is a sign of decreasing inequality and strengthening social cohesion between the groups.}

    There is a growing agreement in the literature over the \textit{historical trend in income and wealth inequality}. 
In particular, it is a commonly held view that the economic dimensions of inequality did not change steadily  in the US, having first decreased during and after the Great Depression, while it begun to increase around 1980   (see \citealp{PikettySaez2003},    \citealp{SaezZucman2016}). 
Even though there is still an ongoing debate about the exact shape of the trend, the \textit{stylized  U-curve pattern} itself has not been challenged (see \citealp{Bricker2016}, \citealp{AutenSplinter2022},  \citealp{Geloso2022}). 

By contrast, the \textit{assortative mating literature is divided about the historical trend in Americans' educational marital homophily}.\footnote{As it is noted by  \cite{NaszodiMendonca2023_RACEDU}, there is no consensus in the literature despite the fact that  ``... unlike the studies on income and wealth inequality, the papers in the assortative mating literature do not perform any perilous exercise of patching together data from different sources. Their main input,  the joint educational distribution of couples,  is provided by the statistical offices  `packed up and parceled' ready for analysis. Also, while under-reporting of income and wealth is a general concern of the researchers, under-reporting  of education level  is not.''} \textit{This paper aims to resolve the puzzling discrepancies and clarify the intergenerational dynamics of educational  homophily in the US.}  We achieve this goal  by shedding light on the link among   
		(i) the \textit{dynamics of Americans' marital educational homophily} between 1960 and 2010,     
		(ii) the \textit{choice of the indicator or method} applied to document the trend, and 
		(iii) the \textit{criteria imposed} on the suitable indicators and methods used to document the trends.

	  We make the following points in this paper. 
	First, \textit{the lack of consensus in the assortative mating literature is not surprising because there is no agreement about   
		what indicators to use for characterizing the directly unobservable homophily.} 
	Studies in the literature offer two main  
	approaches aiming to identify the trends. Either the dynamics are identified    \textit{directly with changes in some ordinal or cardinal statistical indicators of homophily}  
	computed from the joint educational distributions of couples in the generations compared,     
	or, the trends are quantified   \textit{indirectly through the effect of changing homophily on the observed intergenerational change in the prevalence of educational homogamy  with a method controlling for changes in some other drivers of the proportion of homogamous couples}, while also {transforming ordinal measures to cardinal measures of homophily}.

 The second point made in our study is that among the papers identifying homophily indirectly from the prevalence of homogamy, there is \textit{a disagreement about how to control for the confounding factors}, i.e.,  
  	changes in the structural determinant of the proportion of homogamous couples, such as the intergenerational variations in the education level of marriageable young adults. 
  	Also, among the papers identifying homophily directly, there is \textit{a disagreement about what indicators are robust to changes in the structural factor.}  
  	Our second point is the same as that made by \cite{Rosenfeld2008} long before us.\footnote{\cite{Rosenfeld2008} argues that ``One reason the literature
  	on educational assortative mating has produced divergent results is that
  	the changes over time in educational assortative mating are fairly subtle.
  	The underlying marginal distribution of education for both men and women
  	has changed dramatically since the early 20th century, and evaluations
  	of educational assortative mating depend to a great extent on how the
  	rapidly changing educational attainments are controlled for.''} 

As a third point, we stress that \textit{there is no consensus over the set of analytical criteria any indicator has to satisfy to provide a suitable measure of homophily.} 
	Relatedly, we formulate our view reminiscent to the regression argument of Sextus Empiricus 
	 that  
	there is probably no consensus over the criterion fit for deciding which set of analytical criteria to use.\footnote{See the comics  \href{https://www.sites.google.com/site/annanaszodi/funnyphilosophical}{``How the exchange of ideas moves science forward?''} inspired by Monty Python's Witch Burning Trial.}   
	
Finally, to avoid the endless quest of searching for suitable analytical criteria supporting some other criteria, we propose and apply an \textit{empirical criterion} on the direct and the indirect measures of educational homophily.  
Our empirical criterion is     
inspired by the forming consensus in the income and wealth inequality literature:  
\textit{suitable measures of educational homophily should exhibit a robust U-shaped intergenerational trend in the US for the generations born during and after the Great Depression}.  
 
The rest of this paper is structured as follows.   
In Sections \ref{sec:measur} and \ref{sec:Acrit}, we review a comprehensive set of homophily measures proposed in the literature  and an equally comprehensive set of analytical criteria used.   
Our literature review in Section \ref{sec:lit} highlights the link between the homophily measure applied and the identified historical trend in homophily.  
By combining the findings in Sections \ref{sec:measur}, \ref{sec:Acrit} and \ref{sec:lit}, one can gain insight into the extent to which the choice of the criteria imposed on the measures determine the trends in homophily obtained with the measures.  Section \ref{sec:Ecrit} is particularly important: here, we show   
 that one of the  indicators and one of the  methods among those analytical tools reviewed   
perform much better empirically than their alternatives. This finding allows us to select not only the measures, but also the analytical criteria 
reasonable  
to be imposed on the measures.      
Finally, Section \ref{sec:concl} concludes the paper.

\section{Measuring educational homophily}\label{sec:measur}

In this section, we review a set of cardinal and ordinal \textit{homophily indicators}   
that are candidates for characterizing  the strength of educational homophily.   
In addition, we also visit a set of \textit{methods} put forward in the literature for identifying intergenerational changes in the strength of homophily.

\subsection{Measuring homophily directly with statistical indicators}\label{sec:measur_ind}

In this subsection, we review 10 statistical indicators that we define by their closed form formulas. 
In the simplest case of dichotomous assorted trait, i.e., 
where the education level can be either low (L), or  high (H), the joint educational distribution of couples together with the educational distributions of single women and single men  are represented by   the following  contingency table:\\ 
\begin{table}[ht]
	\begin{center}
		\caption{Contingency table with two education levels and singles}
		\begin{tabular}{lll c c c c}  \hline \hline
			$Q^{\text{with singles}}$	& &   & \multicolumn{2}{c}{Women}  &  &         \\ 
			&   & & \multicolumn{2}{c}{in couple} &  &       single \\  \cline{4-5}
			&  &  &                          L  & H & sum  &  men \\ \cline{3-7} 
			\parbox[t]{2mm}{\multirow{2}{*}{\rotatebox[origin=c]{90}{Men}}}& \parbox[t]{2mm}{\multirow{2}{*}{\rotatebox[origin=c]{90}{in c.}}} &L  & $a$     & $b$     & $a+b$  &  $e$ \\ 
			& &                            H & $c$     & $d$     & $c+d$ &  $f$   \\  \cline{3-6}
			& & sum                                  & $a+c$   & $b+d$   & $a+b+c+d$ & \\ 
			& \multicolumn{2}{c}{single women}                              &  $g$    & $h$ &         &  \\   \hline \hline
		\end{tabular}
		\label{tab:CT}
		\vspace{2mm}\\
	\end{center}
\end{table}\\

Whereas the joint educational distribution of couples is given by either of the tables below  
	\begin{minipage}[p]{0.48\columnwidth}
\begin{equation*}\label{eq:Q_noSing}
Q^{\text{2-by-2}}= \begin{bmatrix}
a    & b \\
c   & d
\end{bmatrix}   \;, 
\end{equation*}
\end{minipage}
\hfill
\begin{minipage}[p]{0.48\columnwidth}
\begin{equation*}\label{eq:Q_noSing_nm}
Q^{\text{n-by-m}}= \begin{bmatrix}
N_{1,1} & ...  & N_{1,m} \\
\vdots    &  \vdots       & \vdots \\
N_{n,1}   &...  &N_{n,m}
\end{bmatrix}   \;, 
\end{equation*}

\end{minipage}\\
depending on the number of categories distinguished in the educational trait.

Now, let us review the definitions of the  10 statistical indicators.\\ 
\textbf{(I1)} \textit{Odds-ratio} (it is the most widely used indicator according to \citealp{Chiappori_etal2020}): 
\begin{equation}\label{eq:OR} 
\text{OR}(Q^{\text{2-by-2}})= ad/bc  \;. 
\end{equation}

\textbf{(I2)} \textit{Matrix determinant} (suggested by \citealp{Permanyer2013} and applied by \citealp{Permanyer2019}):
\begin{equation}\label{eq:MD} 
\text{det}(Q^{\text{2-by-2}})= ad-bc  \;. 
\end{equation}

\textbf{(I3)} \textit{Covariance coefficient} (applied by \citealp{Class2017}): 
\begin{equation}\label{eq:cov} 
\text{cov}(Q^{\text{2-by-2}})= \frac{\text{det}(Q^{\text{2-by-2}})}{(a+b+c+d)^2}  \;. 
\end{equation}

\textbf{(I4)} \textit{Correlation coefficient} (applied by \citealp{Kremer1997} and \citealp{FernandezEtAl2005}):
\begin{equation}\label{eq:cor} 
\rho(Q^{\text{2-by-2}})=\frac{\text{det}(Q^{\text{2-by-2}})}{\sqrt{(a+b)(c+d)(a+c)(b+d)}}\;. 
\end{equation}

\textbf{(I5)} \textit{Regression coefficient} (applied by \citealp{Greenwood2014}).  
It is obtained by regressing either the male partners' education on the female partners' education, or the other way around   
\begin{equation}\label{eq:reg} 
\beta_{wm}(Q^{\text{2-by-2}})=\text{det}(Q^{\text{2-by-2}}) / \left[  \left(a+b \right)  \left(c+d \right)   \right] \; \text{or}  
\end{equation} 
\begin{equation*}
\beta_{mw}(Q^{\text{2-by-2}})=\text{det}(Q^{\text{2-by-2}}) / \left[  \left(a+c \right)  \left(b+d \right)   \right]   \;, 
\end{equation*} 
depending on whether the dichotomous variable of wives' education (taking the value of 0 or 1) is explained 
by the dichotomous variable of husbands' education (taking the value of 0 or 1), or vice versa.

\textbf{(I6)} \textit{Aggregate marital sorting parameter} (proposed and applied by  \citealp{Eika2019}): 
\begin{equation*}
\mbox{MSP}(Q^{\text{2-by-2}})= \frac{\mbox{MSP}_L(Q^{\text{2-by-2}}) a +\mbox{MSP}_H(Q^{\text{2-by-2}}) d }{a+d}  \;. 
\end{equation*}
It  is the
weighted average of the marital sorting parameters $\mbox{MSP}_L(Q^{\text{2-by-2}})$ and $\mbox{MSP}_H(Q^{\text{2-by-2}})$. 
Unlike $\mbox{MSP}(Q^{\text{2-by-2}})$, the marital sorting parameters, $\mbox{MSP}_L(Q^{\text{2-by-2}})$ and $\mbox{MSP}_H(Q^{\text{2-by-2}})$, are local measures of sorting. 
The parameter $\mbox{MSP}_L(Q^{\text{2-by-2}})$, corresponding to the L,L couples is  
\begin{equation*}
\mbox{MSP}_L(Q^{\text{2-by-2}})= \frac{{a}/(a+b+c+d)}{a^{\text{counterf.}}/(a^{\text{counterf.}}+b^{\text{counterf.}}+c^{\text{counterf.}}+d^{\text{counterf.}})}  \;,
\end{equation*}
while 
the parameter $\mbox{MSP}_H(Q^{\text{2-by-2}})$, corresponding to the H,H couples is  
\begin{equation*}
\mbox{MSP}_H(Q^{\text{2-by-2}})= \frac{{d}/(a+b+c+d)}{d^{\text{counterf.}}/(a^{\text{counterf.}}+b^{\text{counterf.}}+c^{\text{counterf.}}+d^{\text{counterf.}})}  \;.
	\end{equation*}
The  $\mbox{MSP}_L(Q^{\text{2-by-2}})$ captures the probability that  an L-type man  marries  an L-type woman, relative to
the probability under a counterfactual. Whereas  $\mbox{MSP}_H(Q^{\text{2-by-2}})$ captures the same likelihood ratio, but for the H,H-type couples.

If the counterfactual is the random matching, as it is in the paper by \cite{Eika2019},  then the denominator of $\mbox{MSP}_L(Q^{\text{2-by-2}})$ is $\frac{ (a+b)(a+c) }{(a+b+c+d)^2} $ and 
the denominator of $\mbox{MSP}_H(Q^{\text{2-by-2}})$ is $\frac{ (c+d)(b+d) }{(a+b+c+d)^2}$ making 
$\mbox{MSP}_L(Q^{\text{2-by-2}})= \frac{a (a+b+c+d)} {(a+b)(a+c)}$ and 
$\mbox{MSP}_H(Q^{\text{2-by-2}})= \frac{d (a+b+c+d)} {(c+d)(b+d)}$.
Finally, under the random counterfactual, 
\begin{equation}\label{eq:amsp}
\mbox{MSP}(Q^{\text{2-by-2}})=  \frac{a+b+c+d}{a+d}  \left[\frac{a^2 } {(a+b)(a+c)} + \frac{d^2 } {(c+d)(b+d)}  \right]  \;. 
\end{equation}

\textbf{(I7)} \textit{V-value} (applied by \citealp{Abbott2019}, while they claim to have been proposed first by \citealp{FernandezRogerson2001}): 
\begin{equation}\label{eq:v} 
\text{V}(Q^{\text{2-by-2}})=\frac{\text{det}(Q^{\text{2-by-2}})}{A}\;,
\end{equation} 
where $A=(c+d)(a+c)$ if $b \geq c$ and $A=(b+d)(a+b)$ if $c > b$.

\textbf{(I8)} \textit{Marital surplus matrix} (proposed by \citealp{ChooSiow2006}):
\begin{equation}\label{eq:CS}
\text{MSM}(Q^{\text{with singles}})= \begin{bmatrix}
a/\sqrt{eg}    & b/\sqrt{eh} \\
c/\sqrt{fg}   & d/\sqrt{fh}
\end{bmatrix}   \;. 
\end{equation}

\textbf{(I9)} Scalar-valued \textit{LL-indicator} (proposed by \citealp{LiuLu2006}). 
The formula of the simplified LL-indicator is   
\begin{equation}\label{eq:LL} 
\text{LL}^s(Q^{\text{2-by-2}})=\frac{d - \text{int}(R)  }{\text{min}(b+d, c+d )-\text{int}(R) }\;, 
\end{equation} 
where $R=(c+d)(b+d)/(a+b+c+d)$.
The original LL-indicator is identical to the indicator proposed by Coleman (see Eq. 15 in \citealp{Coleman1958}).  
It is also identical to its  \textit{simplified  version} given by Equation (\ref{eq:LL}) in case sorting is non-negative along the assorted trait in question.   
It is worth to note that the simplified LL-indicator is equivalent to the V-indicator if  $\text{int}(R)=R$ and sorting is non-negative (see Appendix A).

\textbf{(I10)} Matrix-valued  \textit{generalized LL-indicator} (proposed by \citealp{NaszodiMendonca2021}):  
\begin{equation}\label{LiuLugengen}
\text{LL}^{\text{gen}}_{j,k} (Q^{\text{n-by-m}})= 
\text{LL}^s( V_j  Q^{\text{n-by-m}}  W^T_k )    \;,
\end{equation}
where $\text{LL}^{\text{gen}}_{j,k} (Q^{\text{n-by-m}})$ is the  $(j,k)$-th  element of the $\text{LL}^{\text{gen}}$ matrix in case   
of  $Q^{\text{n-by-m}}$ is an $n \times  m$ matrix  with $n \geq 2$, or $m \geq 2$ , or both.   
Further,    
$V_j$ is the $2 \times  n$ matrix \vspace{6mm} \\ 
$V_j = \scriptsize{ \begin{bmatrix}
	\bovermat{\textit{j}}{1    & \cdots &  1} & \bovermat{\textit{n-j}}{ 0  & \cdots & 0}  \\
	0    & \cdots  & 0 & 1  & \cdots  & 1  	
	\end{bmatrix} }$   and  
$W^T_k$ is the $m \times 2$ matrix given by the transpose of \vspace{6mm} \\
$W_k = \scriptsize{ \begin{bmatrix}
	\bovermat{\textit{k}}{1    & \cdots & 1} & \bovermat{\textit{m-k}}{ 0  & \cdots  & 0}  \\
	0    & \cdots  & 0 & 1  & \cdots  & 1  	
	\end{bmatrix} }$ with   $ j \in \{1, \ldots, n-1 \} $, and  $k \in \{1, \ldots, m-1 \}$.

\subsection{Measuring homophily indirectly with counterfactual decompositions}\label{sec:measur_meth}
   
In this subsection, we review a set of \textit{methods}, $\{M_1,...,M_7\}$, that were proposed in the literature for quantifying changes in homophily from an earlier generation to a later generation.

It is a common feature of the methods to be introduced that they \textit{construct a counterfactual contingency table} $Q_{M_i}^{\text{counterf.}}(Q^{\text{late}},Q^{\text{early}})$ from the observed contingency tables $Q^{\text{late}}$ and $Q^{\text{early}}$ characterizing the two generations to be compared. The counterfactual population of couples resembles one of the generations, the earlier born generation, in terms of the structural factor, while it resembles the other generation, the later born generation, in terms of the non-structural factor (see Table \ref{tab:CT2}).   

Once $Q_{M_i}^{\text{counterf.}}(Q^{\text{late}},Q^{\text{early}})$ is constructed with one method or another, it is trivial to  calculate how the prevalence of homogamy would have changed if only the non-structural factor had changed from the earlier generation to the later generation. 
Thereby, one can use the methods to quantify the contribution of the non-structural factor to the change in 
the proportion of homogamous couples. 
We use the contribution calculated, i.e., the ceteris paribus effect of the directly unobservable intergenerational change in homophily, as our \emph{cardinal scalar-valued indirect measure of the change in inequality between different educational groups}.  

After introducing the seven methods proposed in the assortative mating literature, we will make the following salient point in Sections \ref{sec:lit} and \ref{sec:Ecrit}: since   
$Q_{M_i}^{\text{counterf.}}(Q^{\text{late}},Q^{\text{early}})$ depends on the method $M_i$ itself, the quantified change in homophily is also method-specific.          
However, let us first introduce the methods before we turn to their analytical and empirical properties.  

\newpage

\begin{table}[h]
	\caption{Example for the contingency tables $Q^{\text{early}}$ and $Q^{\text{late}}$ representing joint educational distributions of couples in two generations and the counterfactual table $Q_{M_i}^{\text{counterf.}}(Q^{\text{late}},Q^{\text{early}})$ constructed from $Q^{\text{early}}$ and $Q^{\text{late}}$ }
	
	\begin{minipage}[p]{0.48\columnwidth}
		\centering
		\begin{tabular}{llcc c}  
			\hline \hline
			\multirow{2}{*}{$Q^{\text{early}}$} 		& 	 & \multicolumn{2}{c}{Wives in the} &  	\\ 
			& 		 	 & \multicolumn{2}{c}{earlier generation} &  	\\ \cline{3-4}
			&	& $L$  & $H$ & Sum \\ \cline{1-5} 
			\multirow{2}{*}{\rotatebox[origin=c]{90}{Husb.}}	& $L$  & $N^{\text{early}}_{L,L}$  & $N^{\text{early}}_{L,H}$  & $N^{\text{early}}_{L,\cdot}$ \\ 
			& $H$ & $N^{\text{early}}_{H,L}$ & $N^{\text{early}}_{H,H}$   & $N^{\text{early}}_{H,\cdot}$ \\   \cline{2-5}
			& Sum & $N^{\text{early}}_{\cdot,L}$ & $N^{\text{early}}_{\cdot,H}$  & $N^{\text{early}}$ \\ 	 \cline{1-5} \hline 
		\end{tabular}
		
	\end{minipage}
	\hfill
	\begin{minipage}[p]{0.48\columnwidth}
		\centering
		\begin{tabular}{llcc c}  
			\hline \hline
			
			\multirow{2}{*}{$Q^{\text{late}}$} 		& 	 & \multicolumn{2}{c}{Wives in the} &  	\\ 
			& 	         & \multicolumn{2}{c}{later generation} &  	\\ \cline{3-4}
			&	& $L$  & $H$ & Sum \\ \cline{1-5} 
			\multirow{2}{*}{\rotatebox[origin=c]{90}{Husb.}}	& $L$  & $N^{\text{late}}_{L,L}$  & $N^{\text{late}}_{L,H}$  & $N^{\text{late}}_{L,\cdot}$ \\ 
			& $H$ & $N^{\text{late}}_{H,L}$ & $N^{\text{late}}_{H,H}$   & $N^{\text{late}}_{H,\cdot}$ \\   \cline{2-5}
			& Sum & $N^{\text{late}}_{\cdot,L}$ & $N^{\text{late}}_{\cdot,H}$  & $N^{\text{late}}$ \\ 	 \cline{1-5}
			\hline 
			
		\end{tabular}
	\end{minipage}\\

	\begin{center}
		\centering
			\begin{tabular}{llcc c}  
				 \hline
				\multirow{2}{*}{$Q_{M_i}^{\text{counterf.}}(Q^{\text{late}},Q^{\text{early}})$} 		& 	 & \multicolumn{2}{c}{Wives in the} &  	\\ 
				& 		 	 & \multicolumn{2}{c}{hypothetical generation} &  	\\ 
				& 	 & \multicolumn{2}{c}{under the counterfactual} &  	\\ \cline{3-4}
				&	& $L$  & $H$ & Sum \\ \cline{1-5} 
				\multirow{2}{*}{\rotatebox[origin=c]{90}{Husb.}}	& $L$  & $N^{M_i, \text{counterf.}}_{L,L}$  & $N^{M_i, \text{counterf.}}_{L,H}$  & $N^{\text{early}}_{L,\cdot}$ \\ 
				& $H$ & $N^{M_i, \text{counterf.}}_{H,L}$ & $N^{M_i, \text{counterf.}}_{H,H}$   & $N^{\text{early}}_{H,\cdot}$ \\   \cline{2-5}
				& Sum & $N^{\text{early}}_{\cdot,L}$ & $N^{\text{early}}_{\cdot,H}$  & $N^{\text{early}}$ \\ 	 \cline{1-5} \hline \hline
			\end{tabular}
			
		
	\end{center}

	\label{tab:CT2}
\end{table}

The methods  are the\\ 
(M1) \href{https://en.wikipedia.org/wiki/Iterative_proportional_fitting}{Iterative Proportional Fitting} (henceforth IPF) algorithm,\\
(M2) Matrix Determinant-based Approach (henceforth MDbA),\\
(M3) Minimum Euclidean Distance Approach (henceforth MEDA), \\
(M4) \cite{ChooSiow2006} model-based approach (henceforth CSA),\\
(M5) NM-method,\\  
(M6) Generalized NM-method (henceforth GNM), and\\ 
(M7) the method based on the Gale-Shapley matching algorithm (henceforth GS).\\

\textbf{(M1)} The \textit{IPF} algorithm applied to the pair of tables $Q^{\text{early}}$ and $Q^{\text{late}}$  is {defined} by the following two steps to be iterated until convergence.  
First, it factors the rows of  table $Q^{\text{late}}$ 
in order to match the column totals of $Q^{\text{early}}$ (called the target column sums).  
For example, if $Q^{\text{early}}$ and $Q^{\text{late}}$ are defined as in Table \ref{tab:CT2}, then the first step involves multiplying $Q^{\text{late}}$ with the transpose of  $\begin{bmatrix}
N^{\text{early}}_{L,\cdot}/N^{\text{late}}_{L,\cdot} & N^{\text{early}}_{H,\cdot}/N^{\text{late}}_{H,\cdot}    
\end{bmatrix}$. 

The table obtained after the first step (to be denoted by $Q^{'}$) may not have its row totals equal to the row totals of $Q^{\text{early}}$ (called the target row sums).  
In this case, it is necessary to perform a second step.         
As the second step, the IPF factors the columns of $Q^{'}$ to match the corresponding row totals of $Q^{\text{early}}$. 
Again, if $Q^{\text{early}}$ and $Q^{\text{late}}$ are defined as in Table \ref{tab:CT2}, then  the second step involves pre-multiplying $Q^{'}$ with $\begin{bmatrix}
N^{\text{early}}_{\cdot,L}/N^{\text{late}}_{\cdot,L} & N^{\text{early}}_{\cdot,H}/N^{\text{late}}_{\cdot,H}    
\end{bmatrix}$.  

The table obtained  after  this  step (to be denoted by $Q^{''}$) may not have its column totals equal to the target column sums. 
In this case, repeating the first step (which is to be followed by the second step)  is necessary with $Q^{\text{late}}=Q^{''}$.  
Alternatively, we stop the iteration. The table constructed by the IPF is the last $Q^{''}$ table, in case of convergence.  

A notable \textit{feature of the IPF} is that it keeps fixed the non-structural factor (typically referred to as the degree of marital sorting) 
with the \textit{unchanged odds-ratio}, defined by Eq. (\ref{eq:OR}), in case the  assorted trait is dichotomous.
 
The \cite{BS2005} paper is an example for an early application of the IPF in the context of analyzing assortative mating.  
Their approach of constructing  counterfactual joint distributions with the IPF was followed by many researchers (see e.g. \citealp{BreenSalazar2011}, \citealp{HuQian2016},  \citealp{Leesch2022},  \citealp{Shen2021}). Moreover, it was a widely shared view until recently that the IPF is a well-established method for constructing counterfactuals. 

\textbf{(M2)} The \textit{MDbA}, similarly to the IPF, is also a table-transformation method. However, it controls for the non-structural factor with the unchanged \textit{scalar-valued matrix determinant} (see Eq. \ref{eq:MD}), rather then with the odds-ratio. 

The MDbA was proposed by \cite{Permanyer2013} and applied by \cite{Permanyer2019} under the assumption that the assorted trait is dichotomous.  
Since the determinant is defined only for square tables, the MDbA can be applied only if we use the same number of categories for the wives' and the husbands' education level.  
In addition, both the wives' and the husbands' assorted trait has to be dichotomous, otherwise there are multiple tables with the preset target row sums and column sums, and the target value of the determinant.  

\textbf{(M3)} The \textit{MEDA} is {defined} as the table-transformation method that keeps the non-structural factor fixed with the unchanged \textit{scalar-valued V-indicator} (see Eq. \ref{eq:v}). 
The V-indicator interprets as the weight minimizing the Euclidean distance between the table to be transformed (e.g. $Q^{\text{late}}$) and the matrix obtained as the convex combination of the two extreme cases of random and perfectly assortative matching (PAM). More precisely, it is the convex combination of the \textit{expected joint educational distribution of couples under random matching}  and the \textit{joint educational distribution obtained as the outcome of a PAM}, where people in a certain education group can marry someone with a lower education level than their own only if no one in the opposite sex remained unmatched with equal or higher level of education.

\textbf{(M4)} The \textit{CSA} was originally defined with a structural micro founded model  (see \citealp{ChooSiow2006}). However, it can also be  
{defined} in a reduced-form. According to the latter definition, the CSA is the table-transformation method that controls for the non-structural factor 
by keeping fixed the \textit{marital surplus matrix} (see Eq. \ref{eq:CS}).

\textbf{(M5)} The \textit{NM-method} is {defined} as the table-transformation method, where the non-structural factor  
 is controlled for by the unchanged \textit{(generalized) LL-indicator} (see Equations \ref{eq:LL} and \ref{LiuLugengen}).  
It was proposed by \cite{NaszodiMendonca2021} and first applied in a policy brief by \cite{NaszodiPB2019}.

\textbf{(M6)} The \textit{GNM-approach} controls for the aggregate marital preferences over multiple traits  by keeping fixed the \textit{(generalized) trait-specific LL-indicators}. In particular, if the assorted traits considered are the spousal education level and race, then the GNM-approach keeps fixed both the education-specific GLL-indicator and the 
race-specific LL-indicator. The GNM-approach was developed and applied by \cite{NaszodiMendonca2023_RACEDU}. 

\textbf{(M7)} Finally, the \textit{GS-approach} was also proposed to construct counterfactual tables. It assumes that couples are formed with the Gale--Shapley matching algorithm. 
The GS-approach keeps  the non-structural factor, i.e., the aggregate marital preferences over a single dimensional trait, fixed with the \textit{unchanged gender-specific and assorted trait-specific distributions of the reservation points}. This approach was applied by \cite{NaszodiMendonca2019} in the context of analyzing couple formation along education. They used the search criteria of dating site users as a proxy of the reservation points.

\section{A comprehensive set of analytical criteria}\label{sec:Acrit}

An indicator is considered to be a suitable measure, and denoted by $\text{SMH}(Q)$,  if it meets certain criteria. 
The criteria to be imposed on the suitable measures can be chosen from  the following comprehensive set of \textit{analytical criteria}:\\
\textbf{(AC1)} being a \textit{cardinal} measure.\\
\textbf{(AC2)} being \textit{scale invariant} corresponding to the invariance to the change in the total number of couples, while the couples' joint educational distribution is unchanged. Formally, $\text{SMH}$ is scale invariant if  $\text{SMH}(Q)=\text{SMH}(rQ)$ for any $r \in \mathbb{R^+}$.\\
\textbf{(AC3)} \textit{Gender symmetry}, i.e, invariance to interchanging wives' data and husbands' data. 
Formally, $\text{SMH}$ is gender symmetric if $\text{SMH}(Q)=\text{SMH}(Q^T)$, where $Q^T$ denotes the transpose of $Q$.\\
\textbf{(AC4)} \textit{Category symmetry}, defined for the special case of distinguishing only two ordered educational categories of wives and husbands. It is a criterion on the invariance to interchanging the low and the high categories both for wives and husbands. For $\text{SMH}(Q)$, where $Q=Q^{\text{2-by-2}}$, the criterion is defined as  
$\text{SMH} \left(\begin{bmatrix}
a    & b \\
c   & d
\end{bmatrix}  \right) =\text{SMH} \left(\begin{bmatrix}
d    & c \\
b   & a
\end{bmatrix}  \right)$.\\
\textbf{(AC5)}  \textit{Immunity to a certain class of changes in the marginal distributions}.  
It may mean immunity to the type-1 changes in the marginals (see AC5.1), or immunity to the type-2 changes in the marginals (see AC5.2), or immunity to any other type of changes in the marginals (see AC5.3).

$\;\;$\textbf{(AC5.1)} Immunity to the type-1 changes is defined as 
\begin{equation}\label{eq:AC5.1}
\text{SMH}(Q) =\text{SMH} \left(\begin{bmatrix}
 a    &  b \\
\alpha c   & \alpha d
\end{bmatrix}  \right)=
\text{SMH} \left(\begin{bmatrix}
 a    & \alpha b \\
 c   & \alpha d
\end{bmatrix}  \right) \;,
\end{equation} 
where  $Q=Q^{\text{2-by-2}}$ and $\alpha \in \mathbb{R^+}$.
{A type-1 change with $\alpha>1$ can be generated in practice with \textit{active migration policy} used for increasing the number of high skilled individuals in a society.}

$\;\;$\textbf{(AC5.2)}  Immunity to the type-2 changes  is defined as
\begin{equation}\label{eq:AC5.2}
\text{SMH} \left(\begin{bmatrix}
a    & b \\
c   & d
\end{bmatrix}  \right) =\text{SMH} \left(\begin{bmatrix}
(1-\alpha) a    & (1-\alpha) b \\
c +  \alpha a  & d + \alpha b
\end{bmatrix}  \right) =
\text{SMH} \left(\begin{bmatrix}
(1-\alpha) a    &  b+\alpha a \\
(1-\alpha) c  & d + \alpha c
\end{bmatrix}  \right) \;.
\end{equation} \\
Under the type-2 changes in the marginals, $\alpha$ share of the L-type (wo)men are \textit{reclassified} to H-type.
 {A well-known example for a type-2 change with $1>\alpha>0$ is the \textit{educational expansion}.} 
 
 $\;\;$\textbf{(AC5.3)} Immunity to changes in the marginals other than a type-1  or type-2 change.  \\
\textbf{(AC6)} The \textit{weak criterion related to PAM} is defined as follows. If everybody ``marries his or her own type'', the indicator should take its maximum value. 
To formalize this criterion, first we repeat the definition of PAM.   
A matching is a PAM, if  people in a certain education group can marry someone with a lower education level than their own only if no one in the opposite sex remained unmatched with equal or higher level of education.

Now, the definition of the weak criterion itself is that under this criterion, the indicator takes its maximum value in case of PAM with no intermarriages. 
Formally,\\  
$\text{SMH}(Q^{\text{n-by-m}}|Q^{\text{n-by-m}} \text{ is diagonal}, n=m) = max_{P \in \{P_1,..., P_k \}}\text{SMH}(P)$, where $\{P_1,..., P_k \}$ is the set of all possible matches with the same pair of marginals as that of $Q^{\text{n-by-m}}$.  

It is worth to note two things. First, $Q^{\text{n-by-m}}$ is not necessarily a 2-by-2 table, but it can be of any size as long as $n=m$. 
Second, the weak criterion implicitly assumes that the educational distribution of marriageable men (i.e., the column  sums of $Q^{\text{n-by-m}}$) is identical to the educational distribution of marriageable women (i.e., the row sums of $Q^{\text{n-by-m}}$), which is a very special case.\\
\textbf{(AC7)} The \textit{strong criterion related to PAM} is defined as follows. In case of PAM, the indicator takes its maximum value irrespective of the number of intermarriages.
It can be formulated as 
$\text{SMH}(Q^{\text{n-by-m}}|Q^{\text{n-by-m}} \text{ represents a PAM}) = max_{P \in \{P_1,..., P_k \}}\text{SMH}(P)$, where $\{P_1,..., P_k \}$ is the set of all possible matches with the same pair of marginals as that of $Q^{\text{n-by-m}}$.\\ 
\textbf{(AC8)} Our next set of criteria are   \textit{monotonicity} criteria.  
Depending on what variable we impose $\text{SMH}(Q)$  to be monotonous in, we can define various,  
 remarkably different {monotonicity} criteria.

$\;\;$\textbf{(AC8.1)} \textit{Monotonicity in the diagonal cells} is defined as $\text{SMH}(Q^{\text{n-by-n}})\leq \text{SMH}(Q^{\text{n-by-n}}+A)$ for any  positive diagonal matrix $A$ of the same size as $Q^{\text{n-by-n}}$. 

$\;\;$\textbf{(AC8.2)} \textit{Monotonicity in Inter-Generational Mobility} (henceforth IGM) is defined as follows. For any $P$ and $Q$ marital contingency tables representing the joint  distributions of ascribed and attained traits of couples, if the attained traits are more dependent on the ascribed traits for both men and women in the population characterized by $P$ relative to the population characterized by $Q$, 
then  $\text{SMH}(Q)\leq \text{SMH}(P)$.\footnote{While \cite{ClarkCummins2022} derive that the ``greater is assortment, the lower will be social mobility rates'' in a simple model, and a number of empirical papers document this tendency for many countries, we consider it as a criterion reasonable  to be imposed on the $\text{SMH}$s.}  

$\;\;$\textbf{(AC8.3)} \textit{Monotonicity in the number of voluntary singles}, where voluntary singles refer to single individuals to whom not even the most highly educated individuals with  opposite sex are acceptable as partners (see \citealp{NaszodiMendonca2019}). Using the notation in Table \ref{tab:CT}, we can formalize this version of the monotonicity criterion as $\text{SMH}(Q^{\text{with singles}})\leq \text{SMH}(Q^{\text{with singles}}+S)$ for any non-negative matrix $S$ of the form
$\begin{bmatrix}
0    & 0 & e_\text{VS} \\
0   & 0  & f_\text{VS} \\
g_\text{VS}   & h_\text{VS}     \\
\end{bmatrix} $
where  the non-zero elements denote the number of additional voluntary single people, who are either low educated men ($e_\text{VS}$), or   high educated men ($f_\text{VS}$), or low educated women ($g_\text{VS}$), or  high educated women ($h_\text{VS}$) with  $e_\text{VS},f_\text{VS}, g_\text{VS}, h_\text{VS} \in \mathbb{N}$.\\
\textbf{(AC9)} \textit{Immunity to the additional number of involuntary singles}, where additional involuntary singles  are those single individuals in a given generation who have the same marital preferences as those who managed to form couples in an earlier generation. The involuntary singles would also like to be matched and could have been in a couple provided the structural availability of potential partners and competitors had not changed relative to the earlier generation (see \citealp{NaszodiMendonca2019}). Using the notation in Table \ref{tab:CT}, we can formalize the related criterion as
$\text{SMH}(Q^{\text{with singles}}) = \text{SMH}(Q^{\text{with singles}}+Z)$ for any  matrix $Z$ of the form
$\begin{bmatrix}
0    & 0 & e_\text{IS} \\
0   & 0  & f_\text{IS} \\
g_\text{IS}   & h_\text{IS}     \\
\end{bmatrix} $,
where  $e_\text{IS}, f_\text{IS}, g_\text{IS}, h_\text{IS} \in \mathbb{N}$ denote the number of additional involuntary singles.\\
\textbf{(AC10)} \textit{Weak Robustness to the number of Educational Categories} distinguished  (henceforth  weak REC). This criterion can be imposed on methods, rather than on statistical indicators. Let us suppose that method $i$ constructs the following  counterfactual contingency table:  
$Q_{M_i}^{\text{counterf.}}(Q^{\text{late}},Q^{\text{early}})$ (where $Q^{\text{late}}$,  $Q^{\text{early}}$  and \\ $Q_{M_i}^{\text{counterf.}}(Q^{\text{late}},Q^{\text{early}})$ are not necessarily  2-by-2 tables, but they can be of any size).    
In this case, the criterion can be formalized as\\ 
$Q_{M_i}^{\text{counterf.}}\left(\text{Merge}(Q^{\text{late}}),\text{Merge}(Q^{\text{early}}) \right)=\text{Merge}\left( Q_{M_i}^{\text{counterf.}}(Q^{\text{late}},Q^{\text{early}})\right)$.
Under this criterion, the operation constructing the counterfactual table commutes with the operation $\text{Merge()}$ of merging neighboring educational categories.\\
\textbf{(AC11)} \textit{Strong Robustness to the number of Educational Categories} distinguished  (henceforth strong REC). 
As it is suggested by the name of this criterion, the strong version of REC can be met only by those methods that meet the weak REC.  
The strong REC is applicable to indirectly defined, cardinal, scalar-valued indicators obtained with a counterfactual constructing method. 
Under strong REC, the  indicator is \textit{immune to the chosen number of educational categories}, where the indicator
$\text{SMH}(Q^{\text{late}},Q^{\text{early}},Q_{M_i}^{\text{counterf.}})$ captures the scalar-valued ceteris paribus contribution of the changing non-structural factor to a scalar-valued statistics, e.g.  the proportion of homogamous couples.  
The criterion can be formalized as\\
$\text{SMH}(Q^{\text{late}},Q^{\text{early}},Q_{M_i}^{\text{counterf.}})=\text{SMH} \left(\text{Merge}(Q^{\text{late}}),\text{Merge}(Q^{\text{early}}),\text{Merge}(Q_{M_i}^{\text{counterf.}}) \right)$.\\
\textbf{(AC12)} \textit{Signaling Impossible Counterfactuals} (henceforth SIC). This criterion can be imposed on any of the methods, although not all the seven methods reviewed in this paper  fulfill SIC.   
Under SIC, the method should signal if it is not able to control for changes in the educational distributions without constructing an impossible counterfactual.\\

The criteria reviewed (AC1-12) are proposed  in either of the following papers: 
\cite{Eika2019}, \cite{NaszodiMendonca2021}, \cite{Chiappori_etal2020}, \cite{NaszodiMendonca2019}, \cite{Naszodi2022_2m}, \cite{Naszodi2023Grad},  \cite{Naszodi2023WP} and \cite{NaszodiMendonca2023_RACEDU}. Moreover, the relevance and significance of imposing each of these criteria are also pointed out by either of the above studies.

Researchers may disagree about which of the criteria to impose. 
 However, it is not debated by them that \textit{some of the criteria reviewed are mutually exclusive}.  
As a consequence,  none of our 10 direct and 7 indirect measures can meet all criteria introduced (see Tables \ref{tab:crit_ind} and \ref{tab:crit_meth}).  

For instance, the odds-ratio (I1) satisfies AC5.1, but violates AC5.2. 
By contrast, the  aggregate marital sorting parameter (I6) meets AC5.2, while it violates AC5.1. 
This point illustrates that \textit{for testing ``immunity to changes in the marginals'', one has to be specific about the kind of change in the marginals to be controlled for.}

\newcommand{\STAB}[1]{\begin{tabular}{@{}c@{}}#1\end{tabular}}



\begin{landscape}																																			
	\begin{table}[!ht]																																			
		\caption{Analytical criteria and the directly computed statistical homophily indicators}																																	
		\small 																																		
		\begin{flushleft}																																																																									
		\begin{tabular}{ l l l l l c  c c c c c c c c c  }																																		
			\hline \hline																																	
			~ 	&	 ~ 	&	 ~ 	&	 ~ 	&	 ~ 	&																	 \multicolumn{10}{c}{Statistical indicators} 		 \\ \cline{6-15} 						
			~ 	&	 ~ 	&	 ~ 	&	 ~ 	&	 ~ 			&			  \rotatebox{90}{Odds-ratio} 	&	 \rotatebox{90}{Matrix determinant} 	&	 \rotatebox{90}{Covariance coef.} 	&	 \rotatebox{90}{Correlation coef.} 	&	 \rotatebox{90}{Regression coef.}  	&	 \rotatebox{90}{Aggregate MSP} 	&	 \rotatebox{90}{V-value} 	&	 \rotatebox{90}{MSM} 	&	 \rotatebox{90}{LL-indicator} 	&	 \rotatebox{90}{GLL-indicator} 	\\	
			\multicolumn{4}{c}{}   			&	 ~  	&			 (I1) 	&	 (I2) 	&	 (I3) 	&	 (I4) 	&	 (I5) 	&	 (I6) 	&	 (I7) 	&	 (I8) 	&	 (I9) 	&	 (I10) 	 \\ \cline{3-15} 	
			\multirow{15}{*}{\STAB{\rotatebox[origin=c]{90}{\underline{$\;\;\;\;\;\;$Analytical criteria$\;\;\;\;\;\;$}}}}  	&	 \multirow{4}{*}{\STAB{\rotatebox[origin=c]{90}{\underline{$\;\;$Basic$\;\;$}}}} 	&	 (AC1) 	&	 Cardinal 	&	 ~ 	&	Y	&	Y	&	Y	&	Y	&	Y	&	Y	&	N	&	N	&	N	&	N	  \\	
			&	  	&	 (AC2) 	&	 \multicolumn{2}{l}{Scale invariance} 			&	Y	&	Y	&	Y	&	Y	&	Y	&	Y	&	Y	&	Y	&	Y	&	Y	  \\	
			&	  	&	 (AC3) 	&	 \multicolumn{2}{l}{Gender symmetry} 			&	Y	&	Y	&	Y	&	Y	&	N	&	Y	&	Y	&	Y	&	Y	&	Y	  \\	
			&	  	&	 (AC4) 	&	 \multicolumn{2}{l}{Category symmetry} 			&	Y	&	Y	&	Y	&	Y	&	Y	&	N	&	Y	&	NA	&	Y	&	N***	  \\	
			&	 \multirow{8}{*}{\STAB{\rotatebox[origin=c]{90}{\underline{$\;\;\;\;\;\;\;\;$Standard$\;\;\;\;\;\;\;\;$}}}} 	&	 (AC5) 	&	 \multicolumn{2}{l}{Changes in marginal distr. is controlled for}   			&	 ~ 	&	 ~ 	&	 ~ 	&	 ~ 	&	 ~ 	&	 ~ 	&	 ~ 	&	 ~ 	&	 ~ 	&	 ~ 	  \\	
			&		&	$\;\;$ (AC5.1) 	&	\multicolumn{2}{l}{$\;\;$in a given way*} 	        		&	Y	&	N	&	N	&	N	&	N	&	N	&	N	&	N	&	N	&	N	  \\	
			&		&	$\;\;$ (AC5.2) 	&	\multicolumn{2}{l}{$\;\;$in another given way**}			&	N	&	N	&	N	&	N	&	N	&	Y	&	N	&	N	&	N	&	N	  \\	
			&		&	$\;\;$ (AC5.3) 	&	\multicolumn{2}{l}{$\;\;$in alternative ways}	    		&	N	&	Y	&	Y	&	Y	&	Y	&	N	&	Y	&	Y	&	Y	&	Y	  \\	
			&	  	&	 (AC6) 	&	 Weak criterion  	&	 \multirow{2}{*}{PAM}  	&	Y	&	Y	&	Y	&	Y	&		&		&	Y	&		&	Y	&	Y	  \\	
			&	  	&	 (AC7) 	&	 Strong criterion  	&	 ~ 	&	Y	&	N	&	N	&	N	&		&		&	Y	&		&	Y	&	Y	  \\	
			
			&	 	&	 (AC8) 	&	 \multicolumn{2}{l}{Monotonicity}  		  	&		&		&		&		&		&		&		&		&		&		 \\	
			
			&	  	&	 $\;\;$(AC8.1) 	&	 \multicolumn{2}{l}{$\;\;$in diagonal cells} &	Y	&	Y	&	Y	&	Y	&	Y	&	Y	&	Y	&	Y	&	Y	&	Y	 \\	
			&	 \multirow{3}{*}{\STAB{\rotatebox[origin=c]{90}{\underline{Adv.}}}}  	&	 $\;\;$(AC8.2) 	&	\multicolumn{2}{l}{$\;\;$in  inter-generational mobility}	 	     &	N	&	N	&	N	&	N	&	N	&	N	&	Y	&		&	Y	&	Y	  \\	
			&	  	&	 $\;\;$(AC8.3)	&	 \multicolumn{2}{l}{$\;\;$in number of VSs}	 &	NA	&	NA	&	NA	&	NA	&	NA	&	NA	&	NA	&	NA****	&	NA	&	NA	 \\	
			&		&	 (AC9) 	&	\multicolumn{2}{l}{Immunity to additional ISs}			&	NA	&	NA	&	NA	&	NA	&	NA	&	NA	&	NA	&	NA****	&	NA	&	NA	  \\   \hline  \hline	
					\end{tabular} \\
	\end{flushleft}																																							
		\textit{Notes}: Y and N abbreviate yes, and no, respectively. NA stands for not applicable. PAM: Perfectly Assortative Matching, IS: Involuntary Singles, VS: Voluntary Singles, MSP: Marital Sorting Parameter, MSM: Marital Surplus Matrix, LL-indicator: Liu--Lu indicator,  GLL-indicator: Generalized Liu--Lu indicator.\\ 
		*  see Eq.(\ref{eq:AC5.1}),\\ 
		**  see Eq.(\ref{eq:AC5.2}).\\ 
		***   if the assorted trait is multinomial, then it should also be ordered, otherwise the GLL is not applicable.\\
		**** computing the Marital Surplus Matrix does not require to distinguish between voluntary singles and involuntary singles. 										
		\label{tab:crit_ind}																																		
	\end{table}																																			
\end{landscape}																																			

Although it may be obvious to some researchers that by ``immunity to changes in the marginals'' one should mean AC5.1 (see \citealp{Chiappori_etal2020} and reference therein), other researchers, such as \cite{Eika2019}, define the concept of ``immunity'' differently. Otherwise, \cite{Eika2019} would not have applied the aggregate marital sorting parameter (I6) that is immune to the type-2 changes (see AC5.2), rather than to the type-1 changes.

Also, Table \ref{tab:crit_ind}   suggests that instead of requiring the suitable statistical indicators to fulfill all criteria reviewed,  
\textit{we can impose only a restricted set of criteria}.    
When selecting the restricted set of criteria, we should keep in mind the following: 
what criteria are selected determines what statistical indicators are considered to be suitable measures of homophily.     

We stress that since some criteria are mutually exclusive (e.g. AC5.1 and AC5.2), so are some indicators (e.g. I1 and I6): 
it is not rational to believe that several indicators (e.g I1 and I6) fit equally well for the purpose of quantifying changes in homophily.

Similarly to the statistical indicators, none of the \textit{methods} fulfill all the criteria reviewed (see Table \ref{tab:crit_meth}).
For example, AC11 is met by neither of the methods.\footnote{In our view,  the criterion of AC11 should not be imposed on the measures.   
	The reason is that the quantified change in homophily will never be independent of the number of educational categories chosen (even if it is measured 
	by a method satisfying the analytical criterion AC10 of weak REC).  
	\cite{Naszodi2023WP} illustrates this point with an example about assortative mating along age:     
	``Imagine that we analyze matching along age (or any other trait described by a continuous variable). A couple is considered as being homogamous if their age difference is below a certain threshold. Provided the number of age categories is chosen to be extremely high with a threshold as low as one minute, the share of homogamous couples is close to zero. In addition, contrary to common sense, the share of homogamous couples is practically unchanged across any pair of consecutive generations under such an extremely granular set of age categories.''} 
However, even if we eliminate AC11, the methods do not seem to perform well in general.   
On the one hand, this finding is not surprising, because each of the first 6 methods rely on one the 10 statistical indicators and the statistical indicators also fail to meet some of the criteria. 

On the other hand, it is worth to remark that some indirect indicators under-perform remarkably, especially relative to their popularity.  
For instance, one may have the \textit{prior view that the commonly applied IPF meets all the relevant analytical criteria},    
otherwise it would not be used by many scholars.

What can update this prior view? 
First, a close look at Table \ref{tab:crit_meth} showing that the IPF  violates the criteria of AC8.2, AC10, and AC12 (For proof, see \citealp{Naszodi2023WP} and \citealp{Naszodi2022_2m}).\footnote{AC10 is violated not only by the IPF, but also by the MDbA, MEDA and the CS (see Table \ref{tab:crit_meth}).}

Second, what may also challenge the prior view is the recognition that researchers do not choose their methods independently of each other.   
For example, the fact that the IPF is the only readily available method in SPSS harmonizes the choices of SPSS users.  
For the same reason, the  ``majority vote''   does not qualify to decide which analytical tools are suitable for quantifying the trend in homophily.

Let us also look at what arguments might potentially support the prior view that the IPF is fit for constructing counterfactuals.    	
First, it would be reasonable to consider the IPF as a suitable method if AC8.2, and AC10 turned out to be dispensible criteria.     
However, their importance is generally accepted.  
As to criterion AC8.2, there is no disagreement  among researchers that intergenerational mobility and marital sorting are just two sides of the same coin:  
 a society where there is a higher chance for a pauper's son to become a prince is not plausible to be found less open to accepting marriages between paupers and princesses compared to other societies. 
 
 As to the criterion of AC10, it is also reasonable that  
researchers want to avoid  
their measure of homophily to be sensitive  -- due to the poor analytical property of the indicator applied --  to the number of educational categories chosen.   
We emphasize that the sensitivity in question is not empirical in nature: regardless of the data analyzed, certain indicators consistently fail to quantify homophily robust to the categories chosen.

\begin{landscape}																																			
	\begin{table}[!ht]																																			
		\caption{Analytical criteria and the indirect measures of homophily}																																	
		\small 																																		
		\begin{flushleft}																																																					
		\begin{tabular}{ l l l l l c  c c c c c c  }																																		
			\hline \hline																																	
			~ 	&	 ~ 	&	 ~ 	&	 ~ 	&	 ~ 	&																	 \multicolumn{7}{c}{Fixed point}		 \\ 						
			~ 	&	 ~ 	&	 ~ 	&	 ~ 	&	 ~ 	&																	 \multicolumn{7}{c}{transformation-based methods}  		 \\ \cline{6-12} 						
			~ 	&	 ~ 	&	 ~ 	&	 ~ 	&	 ~ 			&			 \rotatebox{90}{Odds-ratio based IPF} 	&	 \rotatebox{90}{MDbA} 	&	 \rotatebox{90}{MEDA} 	&	 \rotatebox{90}{CS-model based approach} 	&	 \rotatebox{90}{GLL-indicator based NM} 	&	 \rotatebox{90}{GLL-indicator based GNM} 	&	 \rotatebox{90}{GS-matching}\\ 								
			\multicolumn{4}{c}{}   			&	 ~  	&			 (M1) 	&	 (M2) 	&	 (M3) 	&	 (M4)   	&	 (M5)   	&	 (M6) 	&	 (M7) 	 \\ \cline{3-12} 							
			\multirow{16}{*}{\STAB{\rotatebox[origin=c]{90}{\underline{$\;\;\;\;\;\;$Analytical criteria$\;\;\;\;\;\;$}}}}  	&	 \multirow{4}{*}{\STAB{\rotatebox[origin=c]{90}{\underline{$\;\;$Basic$\;\;$}}}} 	&	 (AC1) 	&	 Cardinal 	&	 ~ 	&	Y	&	Y	&	Y	&	Y	&	Y	&	Y	&	Y	 \\							
			&	  	&	 (AC2) 	&	 \multicolumn{2}{l}{Scale invariance} 			&	Y	&	Y	&	Y	&	Y	&	Y	&	Y	&	Y	 \\							
			&	  	&	 (AC3) 	&	 \multicolumn{2}{l}{Gender symmetry} 			&	Y	&	Y	&	Y	&	Y	&	Y	&	Y	&	Y	 \\							
			&	  	&	 (AC4) 	&	 \multicolumn{2}{l}{Category symmetry} 			&	Y	&	Y	&	Y	&	Y	&	N**	&	N**	&	N**	 \\	
			&	  	&	     	&	   			&		&		&		&		&		&		&		 \\
			&	 \multirow{5}{*}{\STAB{\rotatebox[origin=c]{90}{\underline{$\;\;$Standard$\;\;$}}}} 	&	 (AC5) 	&	 \multicolumn{2}{l}{Changes in marginal distr. is controlled for*}   			&	Y	&	Y	&	Y	&	Y	&	Y	&	Y	&	Y	 \\							
			&	  	&	 (AC6) 	&	 Weak criterion  	&	 \multirow{2}{*}{PAM}  	&		&		&		&		&	Y	&	Y	&	Y	 \\							
			&	  	&	 (AC7) 	&	 Strong criterion  	&	 ~ 	&		&		&		&		&	Y	&	Y	&	Y	 \\							
			&	  	&	 (AC8) 	&	 \multicolumn{2}{l}{Monotonicity} 	&		&		&		&		&		&		&		 \\							
			&		&	 $\;\;$(AC8.1) 	&	 \multicolumn{2}{l}{$\;\;$in diagonal cells} 	&	Y	&	Y	&	Y	&	Y	&	Y	&	Y	&	Y	 \\							
			&	 \multirow{6}{*}{\STAB{\rotatebox[origin=c]{90}{\underline{$\;\;$Advanced$\;\;$}}}}  	&	 $\;\;$(AC8.2) 	&	\multicolumn{2}{l}{$\;\;$in  inter-generational mobility} 	&	N	&	NA	&	NA	&	NA	&	Y	&	Y	&	Y	 \\							
			&	  	&	 $\;\;$(AC8.3)	&	 \multicolumn{2}{l}{$\;\;$in number of VSs} 	&	NA	&	NA	&	NA	&	N	&	NA	&	NA	&	Y	 \\							
			&		&	 (AC9) 	&	\multicolumn{2}{l}{Immunity to additional ISs}			&	NA	&	NA	&	NA	&	N	&	NA	&	NA	&	Y	 \\							
			&	  	&	 (AC10) 	&	 Weak criterion  	&	 \multirow{2}{*}{REC}  	&	N	&	N	&	N	&	N	&	Y	&	Y	&		 \\							
			&	  	&	 (AC11) 	&	 Strong criterion 	&	 ~ 	&	N	&	N	&	N	&	N	&	N	&	N	&	N	 \\  						
			&	  	&	 (AC12) 	&	 \multicolumn{2}{l}{Signaling impossible counterfactuals} 	&	N	&		&		&	N	&	Y	&	Y	&	NA	 \\  \hline   \hline								
		\end{tabular} \\	
	\end{flushleft}																																
		\textit{Notes}: same as under Table \ref{tab:crit_ind}.  REC: Robustness to the number of Educational Categories distinguished, IPF: Iterative Proportional Fitting algorithm, MDbA: Matrix Determinant based Approach, 
		MEDA: Minimum Euclidean Distance Approach, GS-matching: Gale--Shapley matching,   
		GLL-indicator: Generalized Liu--Lu indicator, 
		GNM-method: Generalized NM-method.\\
		*  each method is designed to control for changes in the marginals in one way or another way.\\ 
		** these methods are applicable if the  assorted trait variable is ordered. 																	
		\label{tab:crit_meth}																																		
	\end{table}																																			
\end{landscape}																																			

Second, some researchers see the prior view confirmed by the authority of the inventors of the IPF.  
The IPF is often referred to as the algorithm first proposed by  \cite{StephanDeming1940}. 
However, if the ``authority card'' is accepted to be played out in the debate concerning method selection,  
then it actually questions, rather than supports the view that the IPF is suitable for constructing counterfactual predictions.   
This is because,  \cite{StephanDeming1940} even warned that  their  algorithm is ``not by itself useful for prediction'' (see  p.444).\footnote{Not surprisingly, their quoted words were ignored by those researchers who attempted to play out the ``authority card'' in favor of the IPF. 
	Also, the fact was overlooked by the same researchers  that \cite{StephanDeming1940} illustrated the application of the IPF on a problem different from constructing counterfactual predictions.}

To conclude this section with a constructive finding,   
we highlight that \textit{some indicators and methods do perform well against the analytical criteria}. 
For instance, both the \textit{scalar-valued LL-indicator} and the \textit{matrix-valued GLL-indicator},  as well as the \textit{NM-method},  the \textit{GNM-method}  and the \textit{GS} are recommended to be applied, because all of them    
 meet the monotonicity criterion in intergenerational mobility (AC8.2). 
Also, unlike the IPF,  the NM and the GNM fulfill the weak criterion on REC (AC10). 
Let us see in the next section, how the analytical tools perform empirically.

\section{The homophily measures applied in the literature and the identified historical trends}\label{sec:lit}

This section explores the link between the measures applied and the historical trends in homophily identified for the US in the assortative mating  literature.

Table \ref{tab:lit}  illustrates the heterogeneity of the empirical results obtained with different analytical tools  (methods/ models/ indicators) in a selective set of papers. 
The  papers reviewed all together identify six qualitatively different trends in homophily (see column 2), while  
they apply 10 different analytical tools (see column 4).

It is worth noting that there is no one-to-one correspondence between the qualitative trends obtained and the tools applied.     
The reason is that besides the selection of the method itself, some other choices of the researchers may also influence the identified trend in homophily:  
e.g., the time period analyzed (see column 5);  
the number of educational categories distinguished (see column 6); 
the choice of the age group of couples observed and the frequency of data studied determining whether each couple is observed only once or in multiple years (see column 7); 
whether the generations compared directly are consecutive or distant (see column 8);
the choice of the decomposition scheme determining whether the interaction between the structural factor and the non-structural factor is controlled for (see column 9).

However, it is not debated in the literature that the choice of the method is crucial:   
even studies 
that do not differ from each other in terms of the data used  and the  decomposition scheme applied, but use different analytical tools,  find diverse trends typically (see e.g. \citealp{Eika2019}, \citealp{LiuLu2006},  \citealp{Naszodi2023Grad}). 
The remarkable  sensitivity of the trend identified to the analytical tool applied will be illustrated in the next section in a context, 
where different tools are applied to the same data using the same decomposition scheme. 

\begin{landscape}																																			

\begin{table}[]
		\caption{Literature review}																				
	\begin{tabular}{rllllccccc} \hline \hline
		& Study                         & Trend in    & Turning                          & Method/       &   \multicolumn{5}{c}{Further aspects}               \\  \cline{6-10}
	    &                               & homophily    & point or                     & model/        &  Period & Num.                & Overl. & Cons.   &With\\
		&                               & (U:U-shaped & period             & indicator     &analyzed &        of     & obs.                 & gen.&    inter-             \\ 
		&                               & NU: not U)  & if                       &               &         & edu.             &             & (if not         &action   \\  
	    &                               &             &  U-shaped                  &               &         & cat.                       &  &  overl.)            &term   \\  
	    & (1)                           & (2)         &  (3)                  & (4)           & (5)     &  (6)                   &   (7)       &     (8)     & (9) \\  \hline

		1&\cite{BreenSalazar2011}       & NU:--        &NA                       & IPF           & 1975-2006& 5                     & Y           & NA            & N     \\
		2&\cite{Chiappori_etal2020}     & NU:+        &NA                       & GCS model     & 1962-2019& 4                     & N           & N           &N \\
		3&\cite{DupuyWeber2018}         & NU:+        &NA                       & CS model      & 1962-2017& 4                     & Y           & NA            & N   \\
		4&\cite{Eika2019}               & NU:+0       &NA                       & Aggregate MSP & 1962-2013& 4                     & N           &N            & N    \\
		5&\cite{Greenwood2014}          & NU:+        &NA                       & Regression    & 1960-2005& >16*                  & Y           &NA           &N  \\
		6&\cite{LiuLu2006}              & NU:+--       &NA                       & LL-indicator  & 1940-2000& 2                     & N           & Y           &N  \\
		7&\cite{Mare1991}               & NU:+0       &NA                       & Odds ratios   & 1930-1987& 5                     & N           & Y           &N  \\
	    8&\cite{Naszodi2021Scheme}      & U:--+        &1990                    & NM-method     & 1980-2010& 3                     & N           & Y           &Y  \\	
	   9&\cite{Naszodi2023Grad}        & U:--+--       & 1990-2000              & NM-method     & 1960-2015& 4                     & N           & Y           &Y  \\
	   10&\cite{NaszodiMendonca2021}    & U:--+        & 1990                    & NM-method     & 1980-2010& 3                     & N           & Y           &Y  \\
	   11&\cite{NaszodiMendonca2023_RACEDU}& U:--+     & 1990                 & GNM-method    & 1960-2010& 3                     & N           & Y           &Y  \\		
	   12&\cite{Permanyer2019}          & NU:+        &NA                       & MDbA   & 1960-2000& 2                     & N           & Y           &N  \\
	   13&\cite{QianPreston1993}        & NU:+        &NA                       & Regression    & 1972-1987& 3                     & N           & Y           &Y  \\
	   14&\cite{Rosenfeld2008}          & NU:--        &NA                       & Odds ratios   & 1940-2005& 5                     & N           & Y           &N  \\
       15&\cite{SchwartzMare2005}       & U:--+        &1960                   & Odds ratios   & 1940-2003& 5                     & Y           & NA           &N  \\
	   16&\cite{Siow2015}               & NU:+        &NA                       & CS model      & 1970-2000& 5                     & N           &N           &N  \\ \hline \hline 
	\end{tabular}
Notes: 
+: increase, --: decrease, 0: stagnant; Y:yes, N: no; 
NA: not applicable; 
*: number of years completed;  
IPF: Iternative Proportional fitting algorithm (also called as the RAS algorithm, or Deming--Stephan algorithm);     
CS model: Choo--Siow model; 
GCS model: Generalized Choo--Siow model (referred to by the authors as the SEV-model);  MSP: marital sorting parameter; 
MDbA: Matrix Determinant-based Approach; 
Num. of edu. cat.: Number of educational categories distinguished; 
Overl. obs.: whether the observations are overlapping; 
Cons. gen.: whether the generations compared are consecutive.

\label{tab:lit}
\end{table}
																																	
\end{landscape}

\section{The empirical criterion on the historical trend}\label{sec:Ecrit}


The  \textit{empirical criterion}  we impose on the suitable statistical indicators and methods is this: 
the analytical tools should identify Americans' marital homophily  
to have a \textit{trend consistent with the income inequality trend}.  
Whether this criterion is fulfilled, will be analyzed in this section both  at the \textit{national level} and the \textit{sub-national level} using census data.   

At the national level, the dynamics of educational homophily should have a U-shaped trend.
More precisely, according to our empirical criterion, 
those analytical tools can be considered to be suitable for the purpose of quantifying the trend in homophily     
that find Americans' preferences over spousal education to have been less and less homophilic in relation to the first four generations born during and after the Great Depression, while the sixth generation should be found by the tools to have had more homophilic preferences than the fifth generation had.

At the sub-national level, the identified state-specific trends in homophily should resemble the corresponding trends in the top 10 percent income share at the state level, which are  also U-shaped for most of the states (see Fig. \ref{fig:Top10} in Appendix B).

The first four generations studied  are the \textit{early Silent generation} (whose members were most active on the marriage market around 1960), the \textit{late Silent generation} (whose members gradually replaced the early Silent generation on the marriage market by 1970), the  \textit{early Boomers} (whose members entered the market around 1980),  and the \textit{late Boomers} (whose members were most active on the market around 1990). 
Whereas the fifth generation is the \textit{early GenX}, whose members were gradually replaced by the sixth generation, the \textit{late GenX} on the market by 2010.

We have \textit{three main motives for imposing our empirical criterion of U-shaped intergenerational trend in Americans' educational homophily}.  
The first is that homophily and economic inequality are  closely related: 
it is a sign of increasing (/decreasing) inequality between different educational  groups if members of a given generation are less (/more) inclined to marry out of their own group relative to their peers in an older generation (\citealp{KatrnakManea2020}).  
Moreover, the \textit{stylized U-pattern of income and wealth inequality} is well-documented in the literature.\footnote{See the introduction about the robust nature of the trend in economic inequality.}    

The second motive is given by the \textit{survey evidence} from the Pew Research Center analyzed  by  \cite{NaszodiMendonca2021} and \cite{Naszodi2023WP}. 
They studied Americans' self-reported preferences over spousal education in four consecutive generations.   
Their results also corroborate the U-shaped pattern of educational homophily at the national level.  

Third, the stylized U-shaped intergenerational trend in educational homophily is also consistent with the   
expansion and the subsequent contraction of the American Social Security System   
that have shaped the relative monetary and non-monetary inequalities in consecutive generations born during and after the Great Depression (see \citealp{Dynarski2003}).  

Having motivated our empirical criterion for the U-shaped trend in homophily at the national level, let us also discuss why we are hesitant to impose the same qualitative trend on each of the US states. 
The \textit{income share of the top 10\% earners}, although exhibited a pronounced \textit{U-shaped trend  over the five decades between 1940 and 1990 at the national level},  
it  \textit{has not had a common trend across all the states}. 
In particular, of the 255 decade-state pairs, only   
228 are consistent with the U-shaped pattern, corresponding to about 90\% of all decade-state pairs.

Another motivation for not imposing the U-shaped homophily trend on all the states is this.  
Which state is assigned to a couple in the census depends on where the couple lived as young adults and not where they grew up, or studied.  
Therefore, a change  in the observed proportion of homogamous couples from one generation to another in some states reflects partly the effects of \textit{non-random domestic and international migration} that is challenging to control for.   
Due to this limitation, we cannot rule out that some uncontrolled effects can make some  decade-state-specific trends in homophily different either  from the U-shaped-consistent pattern, or the state-specific trend in income inequality, or both.   

Third, there are a number of couples who are assigned to neither of the 50 States, nor to District of Columbia.\footnote{In the 1970 census, the number of young couples (with male partners aged between 30 and 34 years) with unknown or unreported residence were more than 800,000.} 
These couples were taken into account in some previous studies documenting the stylized U-shaped historical trend in homophily at the national level (see \citealp{NaszodiMendonca2021}, \citealp{NaszodiMendonca2023_RACEDU}, \citealp{Naszodi2023Grad}). 
However, the same couples are excluded from both of our current analyses using data on sub-national and national levels.     
If couples with certain marital preferences are more likely to belong to the group excluded, it may also effect the homophily dynamics identified for some of the  decade-state pairs analyzed. Similarly to the effect of migration, we cannot control for the potentially present selection bias due to missing records on residence.

In the next subsection,    
we check whether some of the indicators in $\{I_1,...,I_{10}\}$  and some of the methods in $\{M_1,...,M_7\}$    
pass our empirical criterion.

\subsection{Empirical analysis}\label{sec:EMP_FIND}

For the empirical analysis, we use \textit{decennial census data from 6 waves between 1960 and 2010}. 
The data are from IPUMS (see \citealp{IPUMSUSA}). 
The IPUMS data cover the education level of young couples from all   
the 50 US states and the District of Columbia.  

If no observation was missing, then our data would consist of 255 decade-state pairs in total.    
However, the maximum number of decade-state pairs we can work with is slightly lower, being 240.\footnote{No census data are available for Delaware, Idaho,  Montana, North Dakota, South Dakota, Vermont, and Wyoming  for 1970 making it impossible to analyze the trends for the 1960s and 1970s. In addition, we had to eliminate the data for a few state-decade pairs 
	 if  not having any couple of a certain type in the census.}

In our analysis, we distinguish between  three \textit{education levels} at most:  
(i) no high school degree, (ii) high school degree (with no tertiary educational diploma), (iii) at least a college degree.  
However, when working with  indicators defined for dichotomous traits, we use the ``high school graduates-- high school drop outs'' and the  ``college graduates-- no-college degree'' categories. 

We consider a couple to be young if the \textit{male partner is  between 30 and 34 years old}.  
Thereby, our data from the census years 1960, 1970, 1980, 1990, 2000 and 2010  cover husbands and heterosexual male partners from the early Silent generation, the late Silent generation, the early Boomer generation, the late Boomer generation, the early GenX and the late GenX, respectively.\footnote{The wives and female partners may not necessarily be from the same generation as their husbands and male partners are.}  An additional consequence of defining young couples with male partners aged between 30 and 34, is that it makes our decennial data \textit{non-overlapping}: no couple is observed in more than one census wave.

Now, let us see, how the analytical tools perform against our empirical criterion. 
Table \ref{tab:crit_emp} shows that \textit{among the directly and indirectly defined indicators, the LL-measure and the NM-method stand out}, respectively:  
 for almost three-fourth of the decade-state pairs analyzed, we obtain a  
\textit{U-shape-consistent change in the LL-indicator}, irrespective of where exactly we look at the educational distribution. 
More precisely, the share of U-pattern-consistent state-decade pairs   is exactly 75\% and 73\% for the LL with segmentation of the marriage market analyzed along the ``college-- no college'' and ``high school-- no high school'' divides, respectively (see $n_U/N$ in Table \ref{tab:crit_emp}).
 If we study  the segmentations of the market along the two divides jointly with the \textit{NM-method}, then we get  $n_U/N$ as high as 84\%.

In case we restrict our analysis to the segmentation of the marriage market along the ``high school-- no high school'' divide, then 
we find the \textit{covariance} (I3) and the \textit{aggregate marital sorting parameter} (I6) to perform even slightly better than the LL-indicator (see the values of $n_U/N$ in the left block of Table \ref{tab:crit_emp}).

   \begin{landscape}

	
	\setlength{\tabcolsep}{2pt}																																						
	\begin{table}[!htb]
		\begin{center}
			
			\caption{The empirical performance of some direct and indirect measures of marital sorting} 
			\label{tab:crit_emp}																																					
			\begin{tabular}{lccccccccccccccccccc}																																			
				\hline	\hline																																	
				&	\multicolumn{7}{c}{L= no high school degree}															&	$\;\;\;\;$	&	\multicolumn{7}{c}{L= no college  degree}												& $\;\;\;\;$  &\multicolumn{3}{c}{3 edu. levels}		          \\
				&	\multicolumn{7}{c}{H= at least high school degree}															&		&	\multicolumn{7}{c}{H= at least college degree}															& &\multicolumn{3}{c}{are distinguished}         \\
				&	  \rotatebox{90}{Odds-ratio} 	&	 \rotatebox{90}{Matrix det.} 	&	 \rotatebox{90}{Covariance coef.} 	&	 \rotatebox{90}{Correlation coef.} 	&	 \rotatebox{90}{Regression coef.}  	&	 \rotatebox{90}{Aggregate MSP} 	&	 \rotatebox{90}{LL-indicator} 			&		&	  \rotatebox{90}{Odds-ratio} 	&	 \rotatebox{90}{Matrix det.} 	&	 \rotatebox{90}{Covariance coef.} 	&	 \rotatebox{90}{Correlation coef.} 	&	 \rotatebox{90}{Regression coef.}  	&	 \rotatebox{90}{Aggregate MSP} 	&	 \rotatebox{90}{LL-indicator} 			&     &	  \rotatebox{90}{OR-based IPF} 	&	 	 \rotatebox{90}{MEDA} 	&	 \rotatebox{90}{GLL-based NM} 		            \\
				&	(I1)	&	(I2)	&	(I3)	&	(I4)	&	(I5)	&	(I6)	&	(I9)			&		
				&	(I1)	&	(I2)    &	(I3)	&	(I4)	&	(I5)	&	(I6)	&	(I9)	& & (M1)	&		(M3)	&	(M5)			         \\ \cmidrule(lr){2-8}  \cmidrule(lr){10-16} \cmidrule(lr){18-20}
				
				$n_\alpha$	&	60	&	51	&	79	&	75	&	76	&	83	&	78			&		&	78	&	35	&	59	&	64	&	78	&	61	&	89		&	&	60	&	92	&	94	          \\
				$n_\omega$	&	66	&	51	&	85	&	85	&	82	&	90	&	84			&		&	71	&	40	&	63	&	60	&	71	&	67	&	81		&	&	63	&	84	&	87	          \\
				$n_s$	&	126	&	102	&	164	&	160	&	158	&	173	&	162			&		&	149	&	75	&	122	&	124	&	149	&	128	&	170	&		&	123	&	176	&	181	          \\
				$n_U$	&	127	&	109	&	185	&	173	&	171	&	188	&	173			&		&	156	&	66	&	121	&	123	&	160	&	127	&	179	&		&	122	&	183	&	202	          \\
				$N_\alpha$	&	121	&	121	&	121	&	121	&	121	&	121	&	121			&		&	121	&	121	&	121	&	121	&	121	&	121	&	121	&		&	121	&	121	&	121	          \\
				$N_\omega$	&	115	&	115	&	115	&	115	&	115	&	115	&	115			&		&	118	&	118	&	118	&	118	&	118	&	118	&	118	&		&	119	&	119	&	119	          \\
				$N$	&	236	&	236	&	236	&	236	&	236	&	236	&	236			&		&	239	&	239	&	239	&	239	&	239	&	239	&	239	&		&	240	&	240	&	240	          \\
				$n_\alpha/N_\alpha$           	&	50\%	&	42\%	&	65\%	&	62\%	&	63\%	&	69\%	&	64\%			&		&	64\%	&	29\%	&	49\%	&	53\%	&	64\%	&	50\%	&	74\%	&		&	50\%	&	76\%	&	78\%	          \\
				$n_\omega/N_\omega$           	&	57\%	&	44\%	&	74\%	&	74\%	&	71\%	&	78\%	&	73\%			&		&	60\%	&	34\%	&	53\%	&	51\%	&	60\%	&	57\%	&	69\%	&		&	53\%	&	71\%	&	73\%	          \\
				$n_s/N$           	&	53\%	&	43\%	&	69\%	&	68\%	&	67\%	&	73\%	&	69\%			&		&	62\%	&	31\%	&	51\%	&	52\%	&	62\%	&	54\%	&	71\%	&		&	51\%	&	73\%	&	75\%	          \\
				$n_U/N$           	&	54\%	&	46\%	&	78\%	&	73\%	&	72\%	&	80\%	&	73\%			&		&	65\%	&	28\%	&	51\%	&	51\%	&	67\%	&	53\%	&	75\%	&		&	51\%	&	76\%	&	84\%	          \\
				
				\hline	\hline																																	
				
			\end{tabular}																											
			
		\end{center}

		\textit{Notes}: For the statistics reported in the left and middle blocks of the table, high education level is defined either as ``at least high school degree'', or as ``at least college degree''. 
		$N$ denotes the total number of decade-state pairs analyzed.  
		$N$ is close to $51\times5=255$, but somewhat lower because of missing data for some states in certain census years and also because no couple with certain level of education  is registered in the census for some state-census year pairs. 
		$N_\alpha$ and $N_\omega$ denote the number of decade-state pairs in the first half (from Alabama to Mississippi) and the second half (from Missouri to Wyoming) of the states ordered alphabetically, respectively. 
 		$n_U$ is calculated as the number of decade-state pairs with \textit{U-shape-consistent direction of change} in the indicator (i.e.,  negative  for the 1960s, 1970s, 1980s, and positive  for the 1990s and 2000s).  
		$n_s$ denotes the number of decade-state pairs, where the identified trend in homophily is consistent with the \textit{state-specific income inequality trend} defined by the top 10 percent income share.   
		$n_\alpha$ and $n_\omega$ denote the number of state-specific income inequality trend-consistent decade-state pairs in the first half and the second half of the states, respectively. 
		$n_U/N$,  $n_s/N$, $n_\alpha/N_\alpha$, and $n_\omega/N_\omega$ quantify the empirical performance of the analytical tools.
		
	\end{table}

\end{landscape}																										

However, both of these indicators perform almost as poorly in analyzing the ``college-- no college'' divide as a random indicator with a 50\% chance of being right about the direction of intergenerational change in homophily (see the middle block of Table \ref{tab:crit_emp}).  In addition, both the average performance  of the \textit{covariance} (I3), calculated as $(185+121)/(236+239)=64\%$,  and that of the \textit{aggregate marital sorting parameter} (I6), calculated as $(188+127)/(236+239)=66\%$, are well below the average performance  of the {LL} (I9), calculated as $(173+179)/(236+239)=74\%$.

Regarding the \textit{robustness of the analytical tools' ranking}, Table \ref{tab:crit_emp} shows that the relative positions do not change if we use the resemblance to the    
state-specific income inequality dynamics  as our criterion (see $n_s/N$ in Table \ref{tab:crit_emp}), rather than the resemblance to the    
U-shaped income inequality trend at the national level. 
    
The {ranking of the analytical tools is also robust} to whether we study  all the 50 states and DC (see $n_s/N$ in Table \ref{tab:crit_emp}), or only the states whose names  are before Missouri in the alphabetical order (see $n_\alpha/N_\alpha$ in Table \ref{tab:crit_emp}), or only the states whose names are  after Mississippi in the alphabet (see $n_\omega/N_\omega$ in Table \ref{tab:crit_emp}). Putting it differently, if we select the best performing tools by using the data only for the \textit{first 25 states from Alabama to Mississippi}, then the \textit{out-of-sample performance} of the selected tools, evaluated by using data of the \textit{other 26 states}, will also be the best. 

In particular, the LL-measure  performs the best among the directly calculated measure  in $\{I_1, I_2, I_3, I_4, I_5, I_6, I_9 \}$  with its average performance indicator taking the value $69\%$, calculated as   $(78+89)/(121+121)$  using data of the \textit{first 25 states}.    
By using data of the \textit{second half of the states}, it is again the LL that triumphs with its average performance indicator taking the value $71\%$  (calculated as   $(84+81)/(115+118)$).   
Among the candidates for the indirect measure of homophily, it is the NM that performs the best not only according to $n_\alpha/N_\alpha$, but also according to $n_\omega/N_\omega$ (see the right block of Table \ref{tab:crit_emp}).

It is worth noting that the \textit{NM outperforms both the IPF and the MEDA} also according to $n_U/N$ and  $n_s/N$ 
   (see the  block of Table \ref{tab:crit_emp} on the right). 
With regard to the \textit{IPF}, it identifies an \textit{increase in homophily} for most of the state-decade pairs (see Fig. \ref{fig:IPF} in Appendix B).  
Also, it finds homophily to have had a  \textit{steady increase at the national level} for the five decades analyzed (see the black line of Fig. \ref{fig:IPF} in Appendix B). 
 
As far as the {\textit{MEDA}} is concerned, the trend it identifies for the US exhibits a \textit{W-shape}, rather than a U-shape (see the black line of Fig. \ref{fig:MEDA} in Appendix B). Moreover, its state-specific dynamics are U-shape-consistent for a fewer number of state-decade pairs 
 than those intergenerational trends obtained with the NM  (see Figures \ref{fig:MEDA}, \ref{fig:NM} in Appendix B). 

Regarding  the \textit{analytical tools not covered by Table \ref{tab:crit_emp}}, some of them would score similarly to the IPF. 
Presumably, the \textit{Marital Surplus Matrix} (I8) and the \textit{CS-model} (M4) are such, because the CS  is also found 
to identify a steady increase or stagnation in homophily at the national level (see  Subfigure \ref{fig:CSGNM}a  
and rows 2, 3, 16 of Table \ref{tab:lit} reporting the qualitative findings in \citealp{Chiappori_etal2020}, \citealp{DupuyWeber2018}, and \citealp{Siow2015}).  

With regard to the \textit{MDbA} (M2), if it was applied to analyze homophily either along the  ``high school-- no high school'' divide, or the ``college-- no college'' divide, then its performance in terms of $n_U/N$ and $n_s/N$ would be no different from that of the matrix determinant-indicator (I2) that  the MDbA is built on. Since the MDbA-transformed table is not unique for multinomial traits, such as  education in our specific problem  with three different levels distinguished, it is not straightforward how to compare the MDbA's  performance to those of some other analytical tools. 

It is worth noting, that some other analytical tools, also not covered by Table \ref{tab:crit_emp},  perform  similarly well as the LL and the NM.  
For instance,  if the  \textit{V-value} (I7) was applied to analyze sorting either along the dichotomous traits of ``high school-- no high school'', or  ``college-- no college'', then its  performance would be equally good as that of the LL due to the near-equivalence of these indicators (see Appendix A). 

Finally, let us turn to the last two tools  omitted from Table \ref{tab:crit_emp}. 
The indicators constructed with the \textit{GNM} (M6),  and the \textit{GS-matching} (M7), similar to the Marital Surplus Matrix (I8) and the indicator defined by the CS (M4), are challenging to be compared to the 10 analytical tools  in Table \ref{tab:crit_emp}, because the former indicators are 
calculated from a  broader set of information.\footnote{\cite{NaszodiMendonca2023_RACEDU} calculate the indicators defined by the GNM from the \textit{joint educational and racial distribution of couples}. The Marital Surplus Matrix and the indicator defined by the CS can be calculated only if the \textit{education level of single individuals} is known on top of the education level of individuals in couples. To apply the GS, \citealp{NaszodiMendonca2019} use information not only on singles, 
	but also the \textit{search criteria of dating site users}.} 
Still, presumably, both the GS-matching,  and the GNM perform just as well as the NM,  because they identify qualitatively the same trend as the NM at the national level  
(see  \citealp{NaszodiMendonca2019},  Subfigure \ref{fig:CSGNM}b,  
    and rows 8, 9, 10, 11 of Table \ref{tab:lit}).  

To sum, our results suggest that it is not impossible to obtain a U-shaped pattern for educational homophily 
with an analytical tool  other than the LL-indicator  and the NM-method.  
However, \textit{out of the 10 analytical tools that can be calculated from couples' education level only,  
it is exclusively the  LL-indicator and the GLL-indicator-based NM-method that document a U-shaped trend robust to where exactly one looks at the educational distribution and whether the analysis is performed at the sub-national level, or the national level}.

\subsection{Discussion}\label{sec:disc}

Sections \ref{sec:Acrit} and \ref{sec:lit} showed that there is a disagreement in the literature on 
(i) what indicators are suitable to characterize homophily,   
(ii) what set of analytical criteria to impose on the suitable indicators, and  
(iii) what was the qualitative intergenerational trend in homophily in the US over the five decades between 1960 and 2010.

The disagreement calls for a selection of the analytical tools,  the criteria,  and the trends. 
It may be tempting to perform the selection by relying either on our intuition, 
or on the authority of some researchers, 
or on a simple ``majority vote''. 

Regarding our intuition, it often fails.\footnote{\cite{kahneman2011thinking} offers a number of funny examples illustrating this point.} 
Other researchers, and ourselves, are not infallible either. 
Finally, the ``majority vote'' is similarly problematic as relying either on others, or our intuition:     
in this paper, we argued that the ``votes'' of  researchers would  not reflect independent views. 
As to a ``majority vote'' among analytical tools, rather than among researchers, we make the following point.     
Even if a large set of indicators suggests that the historical trend in homophily and inequality is of a given shape,   
the size of this set cannot be used as a final argument in favor of  the shape in question.\footnote{In other words, standard robustness analyzes to the choice of the indicator do not qualify as a criterion on the set of criteria.}  
The reason is that these indicators can be from the same family.  
If the entire family of indicators is ill suited for the analysis then none of the indicators in the family is suitable irrespective of the size of the family.\footnote{Consider 
	an analogous example about the distribution of extreme values of financial returns.  
The most impressive theorem in extreme value theory, the Fisher--Tippett--Gnedenko theorem states that 
statistical distribution of the largest value drawn from a sample of a given size has only three possible shapes: 
it is either a Weibull, a Fréchet or a Gumbel distribution depending on the distribution of the population that the sample is taken from.  
If the Gumbel-family is ill suited to model value-at-risk of a financial portfolio then any distribution in this family (including the Gaussian distribution) is inadequate to model extreme returns irrespective of the high number of distributions we can name in this family.} 

We believe that there is a royal road neither to the suitable analytical criteria, nor to the indicators fit for the purpose, nor to the plausible quantitative historical trends.   
Accordingly, we have taken neither the path of imposing certain intuitive analytical criteria that can be endlessly questioned, 
nor applied Ethos as a mode of persuasion,  
nor have taken the path of applying a ``majority vote''.  

Rather,  we imposed a \textit{well-motivated empirical criterion} on the homophily dynamics between 1960 and 2010 in the US. 
Our empirical criterion \textit{selects 5 tools  
from our 16-piece toolbox}: the  scalar-valued LL-indicator, the matrix-valued GLL-indicator, the  NM-method,  the GNM-method,  and the GS.  
The tools selected are also found to perform well against certain analytical criteria. 

Upon acceptance of our empirical criterion, we propose to \textit{select the following criteria into our restricted set of criteria}:  
scale invariance (AC2),  gender symmetry (AC3),  invariance to certain changes in the marginals that are neither the type-1, nor the type-2 changes (AC5.3), the weak criterion related to perfectly assortative matching (AC6), the strong criterion related to perfectly assortative matching (AC7), monotonicity criterion in intergenerational mobility (AC8.2), and the weak criterion on robustness to the number of educational categories distinguished (AC10).

 The significance of our carefully performed indicator selection and method selection is the following.  
Once we have some analytical tools suitable for quantifying the historical trend in   
inequality in the US over the twentieth century, the tools can be applied to other countries and periods as well, for which a definitive narrative is lacking so far. 
 In particular, the NM-method combined with micro level data on who marries whom (and who cohabits with whom) can be used to quantify the  inequality dynamics pertaining to  countries, 
where we have neither any strong prior on the qualitative trends, nor reliable micro data on tax declarations to compute income inequality indicators, nor surveys on individuals self-reported marital preferences.\footnote{See the web-page of the International Demographic Inequality Lab (\href{https://idil.li}{IDIL.LI}) reporting  
	the historical trend in overall inequality in 80 countries calculated from data on couples.} 

\section{Conclusion}\label{sec:concl}

In this paper, we addressed two  questions. First, how to measure changes in educational homophily from one generation to the next with the purpose of identifying the historical trend in inequality between groups of people with different level of education (and with different ability to generate income).
Second, what set of analytical criteria should be used in order to select the  suitable measures of homophily.

Following the non-argumentative approach of the animals in Orwell's  Animal farm, we could have said that 
U-shape is good, non-U-shape is bad at characterizing the historical trend in homophily in the US.  
Similarly, we could have said that those analytical criteria fit for the purpose of selecting suitable homophily measures    
 that are violated by the measures resulting in non-U-shaped trends,    
while those criteria are not fit for the same purpose  that are violated by the analytical tools resulting in U-shaped trends.

In this paper, we applied a somewhat more sophisticated approach.   
First, we reviewed a comprehensive set of analytical tools and criteria. 
Second, we studied which of the tools introduced meet the criteria reviewed.   

Third, we illustrated with a brief literature review the point that what intergenerational trend in  Americans' homophily is identified in the studies is highly sensitive to the analytical tool used, as well as to the choice of the criteria imposed on the suitable analytical tools, along some other choices of the researchers.    

Fourth, we argued in favor of the U-shaped intergenerational trend in Americans' educational homophily by referring to the well-documented U-patterns common across various phenomena related to the revealed educational homophily, such as income inequality, size of the American Social Security System,  and self-reported homophily. 

Finally, we showed that the LL-indicator, and the NM-method do not only have attractive analytical properties, but they represent analytical tools that result in a robust U-shape pattern of revealed homophily in the US between 1960 and 2010.

\bibliography{Bib_ODDS_01}

\newpage

\begin{appendices}
	
	\renewcommand{\theequation}{A\arabic{equation}}
	\renewcommand{\thetable}{A\arabic{table}}
	\renewcommand{\thefigure}{A\arabic{figure}}

	\setcounter{equation}{0}
	\setcounter{table}{0}
	\setcounter{figure}{0}

	\section*{Appendices}\label{sec:Appendices}
	
	\subsection*{Appendix A}\label{sec:Appendix_A}
	In this appendix, we show that the simplified Liu--Lu indicator (LL-indicator) is identical to the V-value under certain conditions. 
	Trivially, under their identity, the MEDA is identical to the NM-method.

	To recall, $\text{V}(Q)=\frac{\text{det}(Q)}{A}$,    
	where $Q=\begin{bmatrix}
	a    &  b \\
	c   & d
	\end{bmatrix}$, and  $A=(c+d)(a+c)$ if $b \geq c$ and $A=(b+d)(a+b)$ if $c > b$ (see Eq. \ref{eq:v} ).  
	The formula of the simplified LL-indicator is  $\text{LL}^s(Q)=\frac{d - \text{int}(R)  }{\text{min}(b+d, c+d )-\text{int}(R) }$, where $R=(c+d)(b+d)/(a+b+c+d)$ (see Eq. \ref{eq:LL}). 
	
	We will prove that if $\text{int}(R)=R$, then $\text{LL}^s(Q)=\text{V}(Q)$. 
	Let us suppose that $b \geq c$.
	Then, $\text{V}(Q)=\frac{ad-bc}{(c+d)(a+c)}$. 
	Furthermore, if $\text{int}(R)=R$ and $b \geq c$, then   $\text{LL}^s(Q)=\frac{d - R }{c+d-R}$. 
	By substituting the definition of   $R=(c+d)(b+d)/(a+b+c+d)$, we get $\text{LL}^s(Q)=\frac{d - (c+d)(b+d)/(a+b+c+d) }{c+d-(c+d)(b+d)/(a+b+c+d)}$.

	So, what needs to be shown is that 
	\begin{equation}\label{eq:veqLL} 
	\frac{ad-bc}{(c+d)(a+c)} = \frac{d - (c+d)(b+d)/(a+b+c+d) }{c+d-(c+d)(b+d)/(a+b+c+d)} \;.
	\end{equation} 
	
	By multiplying both sides of Eq. \ref{eq:veqLL} by  $[c+d-(c+d)(b+d)/(a+b+c+d)](a+b+c+d)(c+d)(a+c)$, we get
	\begin{equation}\label{eq:veqLL1} 
	({ad-bc}) [(c+d) (a+b+c+d) - (c+d)(b+d)]    = [d (a+b+c+d)- (c+d)(b+d) ] (c+d)(a+c)\;.
	\end{equation} 
	
	Since the two sides of  Eq. \ref{eq:veqLL1} are equal, we get that the simplified Liu--Lu indicator is the same as the V-value if 
	$\text{int}(R)=R$. 
	
	Finally, we make the point that unlike the LL, the GLL cannot be equivalent to the V-value, since the GLL is matrix-valued, while the V-value is scalar-valued. Relatedly, the MEDA can be identical to the NM-method unless the assorted trait is dichotomous.

\subsection*{Appendix B}\label{sec:Appendix_B}

\newpage

\begin{landscape}																																			

\begin{figure}
	\begin{center}
		
	\caption{Trends in top 10 percent income share}
	
	\begin{subfigure}{0.6\textwidth}
		\centering
		\includegraphics[width=.9\linewidth]{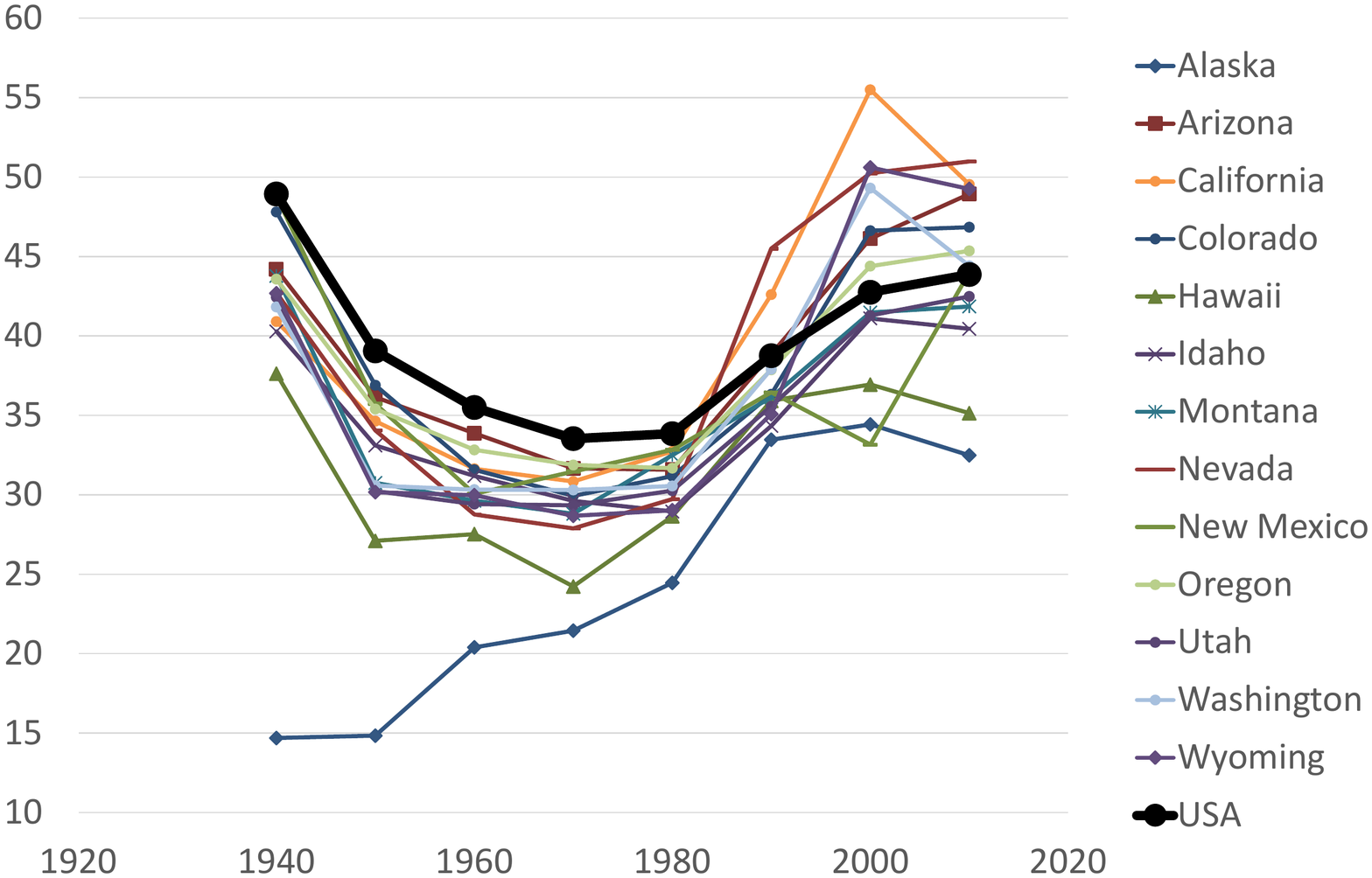}
		\caption{West}
		\label{fig:Top10_W}
	\end{subfigure}%
	\begin{subfigure}{0.6\textwidth}
		\centering
		\includegraphics[width=.9\linewidth]{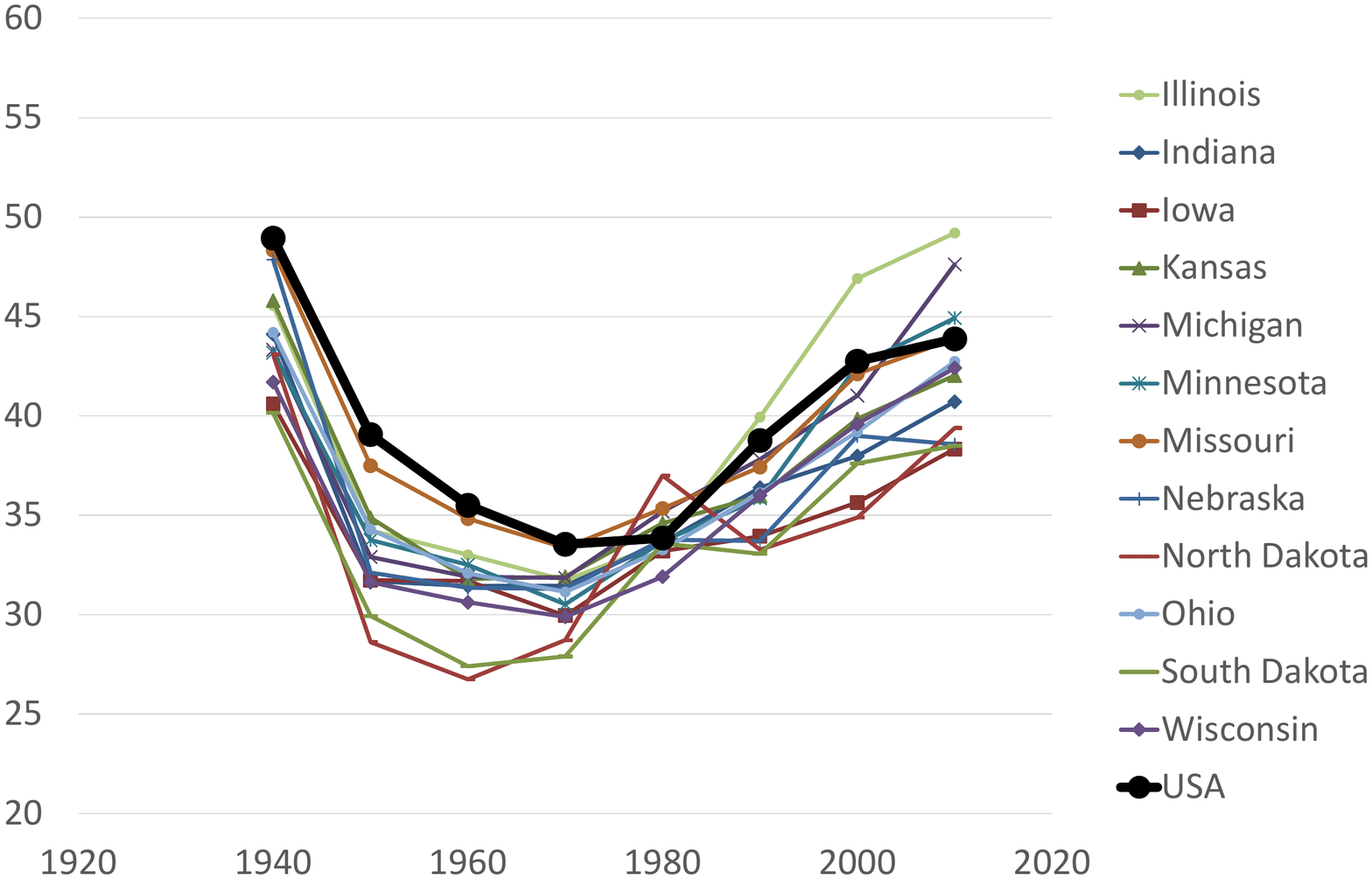}
		\caption{Midwest}
		\label{fig:Top10_MW}
	\end{subfigure}\\
	
	\begin{subfigure}{0.6\textwidth}
		\centering
		\includegraphics[width=.9\linewidth]{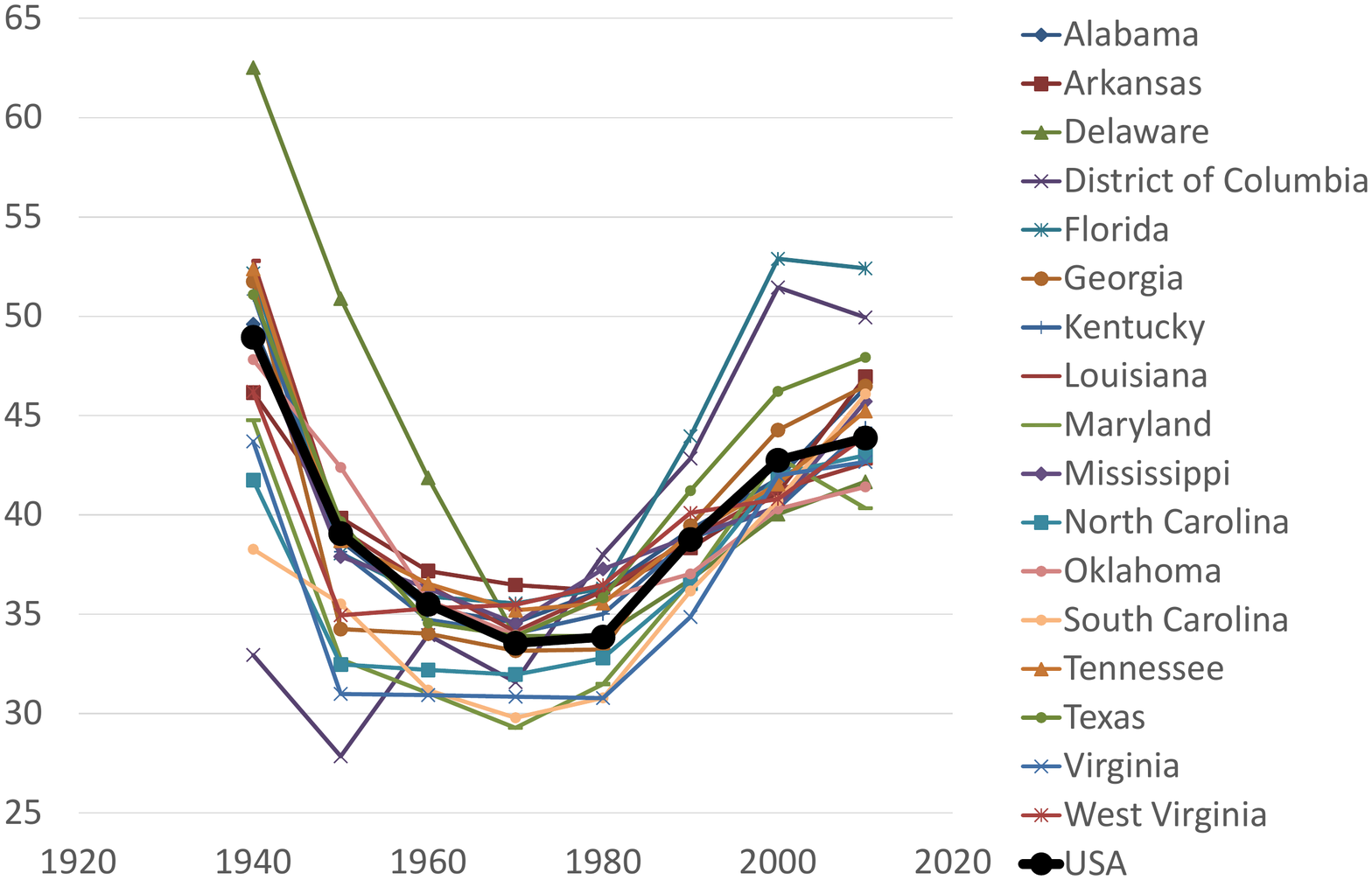}
		\caption{South}
		\label{fig:Top10_S}
	\end{subfigure}%
	\begin{subfigure}{0.6\textwidth}
		\centering
		\includegraphics[width=1\linewidth]{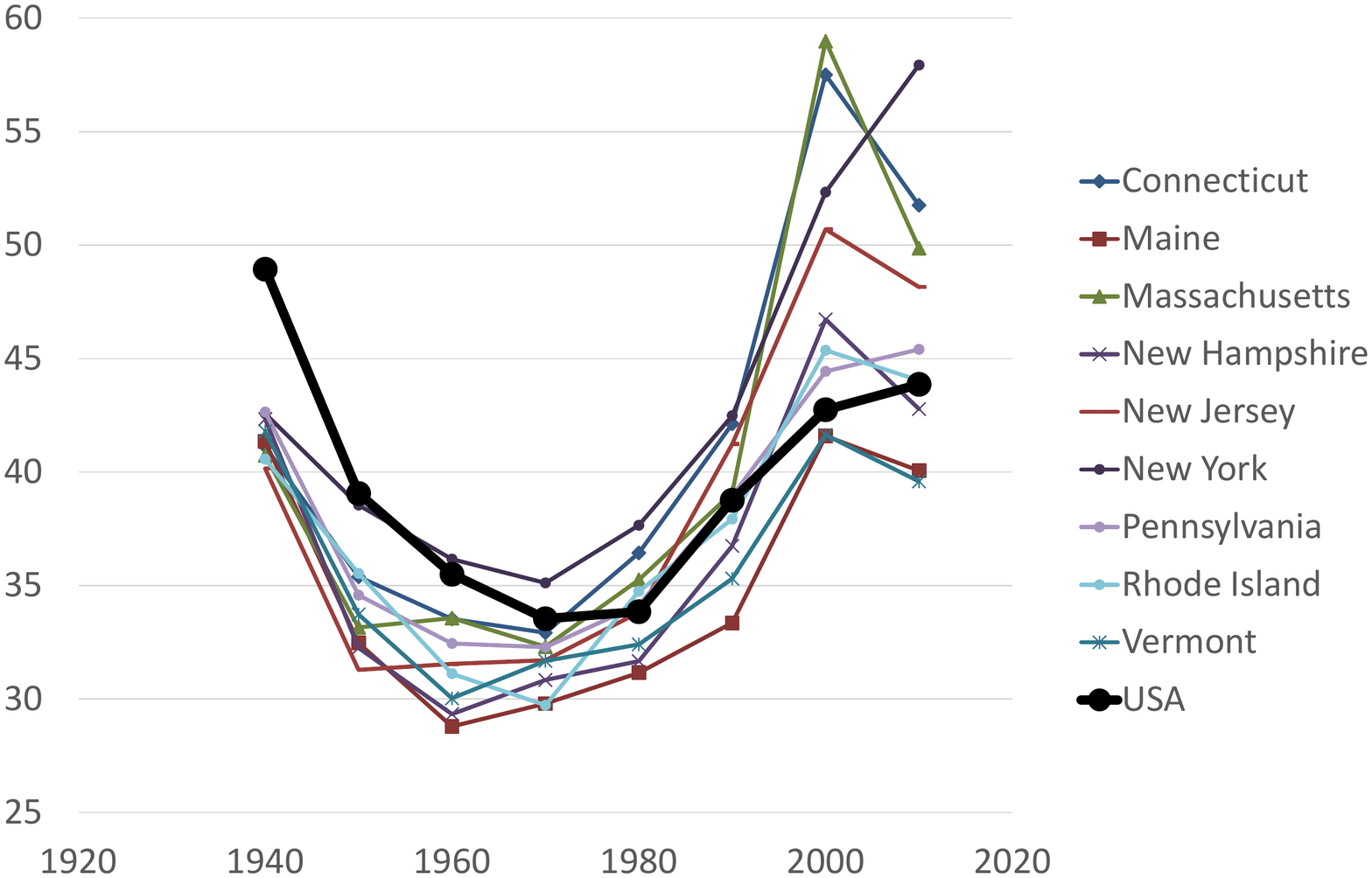}
		\caption{Northeast}
		\label{fig:Top10_NE}
	\end{subfigure}
	
	\label{fig:Top10}
	\end{center}	

	Source: WID.WORLD.

\end{figure}

\end{landscape}

\newpage
\begin{landscape}																																			
\begin{figure}
	\caption{Trends in educational homophily  obtained with the IPF algorithm}
	\begin{center}
		
	\begin{subfigure}{0.7\textwidth}
		\centering
		\includegraphics[width=.9\linewidth]{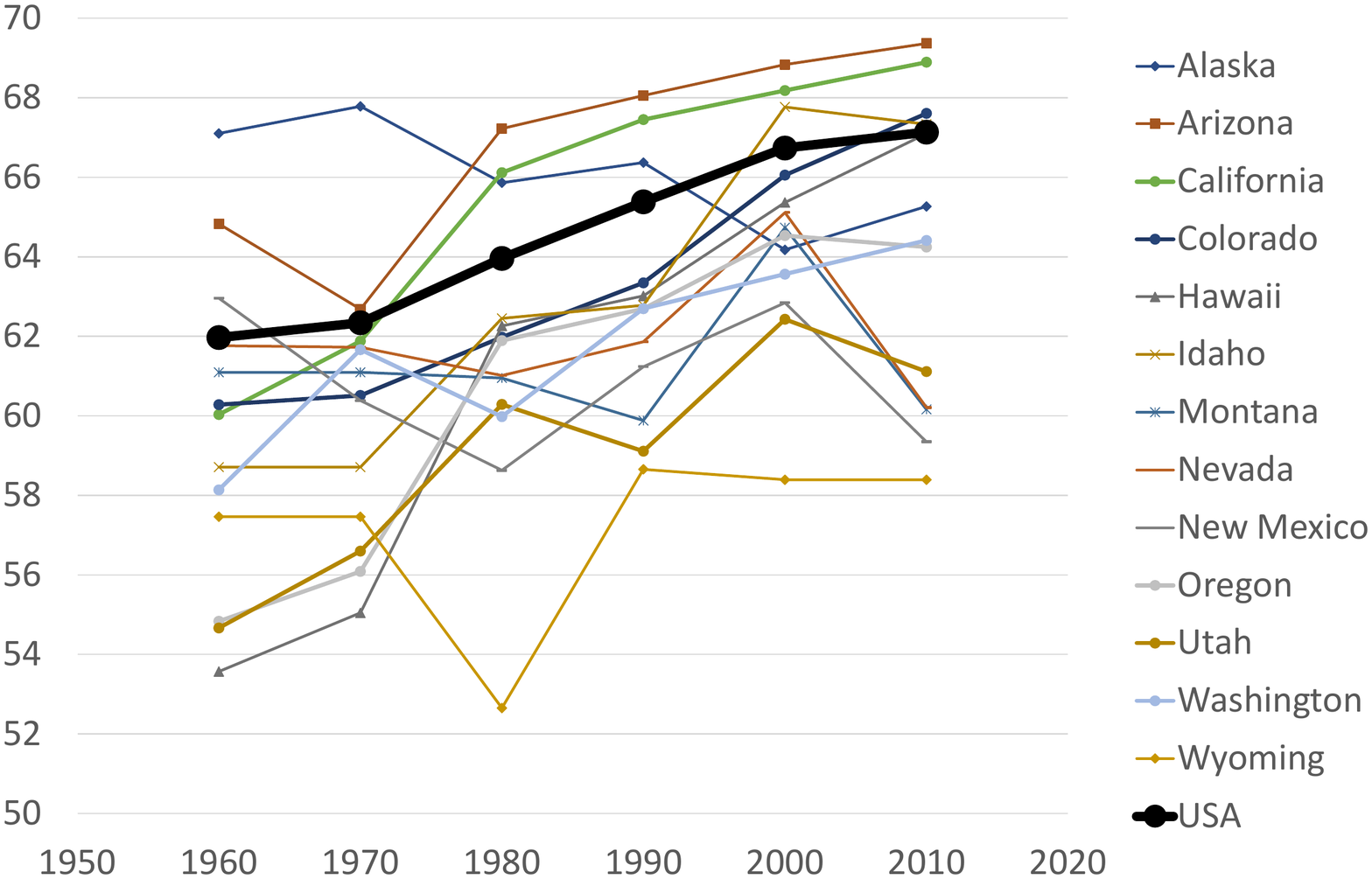}
		\caption{West}
		\label{fig:IPF_W}
	\end{subfigure}%
	\begin{subfigure}{0.7\textwidth}
		\centering
		\includegraphics[width=.9\linewidth]{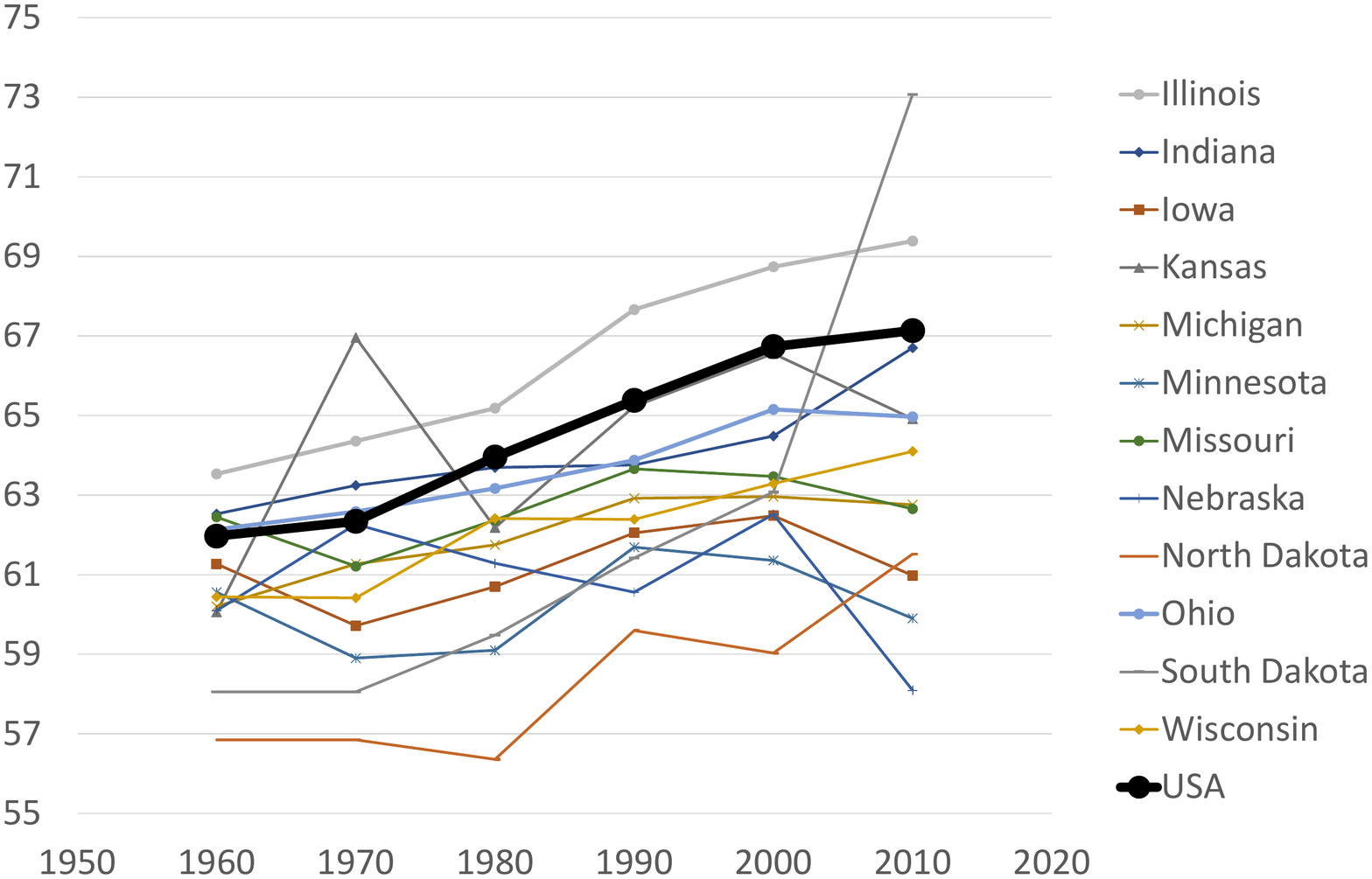}
		\caption{Midwest}
		\label{fig:IPF_MW}
	\end{subfigure}\\
	\label{fig:IPF}
	\end{center}
\textit{Note}: see below Subfigures \label{fig:MEDA}c,d. 

\end{figure}
	\end{landscape}																																			
	\newpage

	\begin{landscape}																																			
\begin{figure}
\begin{center}
	
\textbf{FIGURE A2 (continued)}: Trends in educational homophily  obtained with the IPF algorithm


	\begin{subfigure}{0.7\textwidth}

\addtocounter{subfigure}{+2}
		\centering
		\includegraphics[width=.9\linewidth]{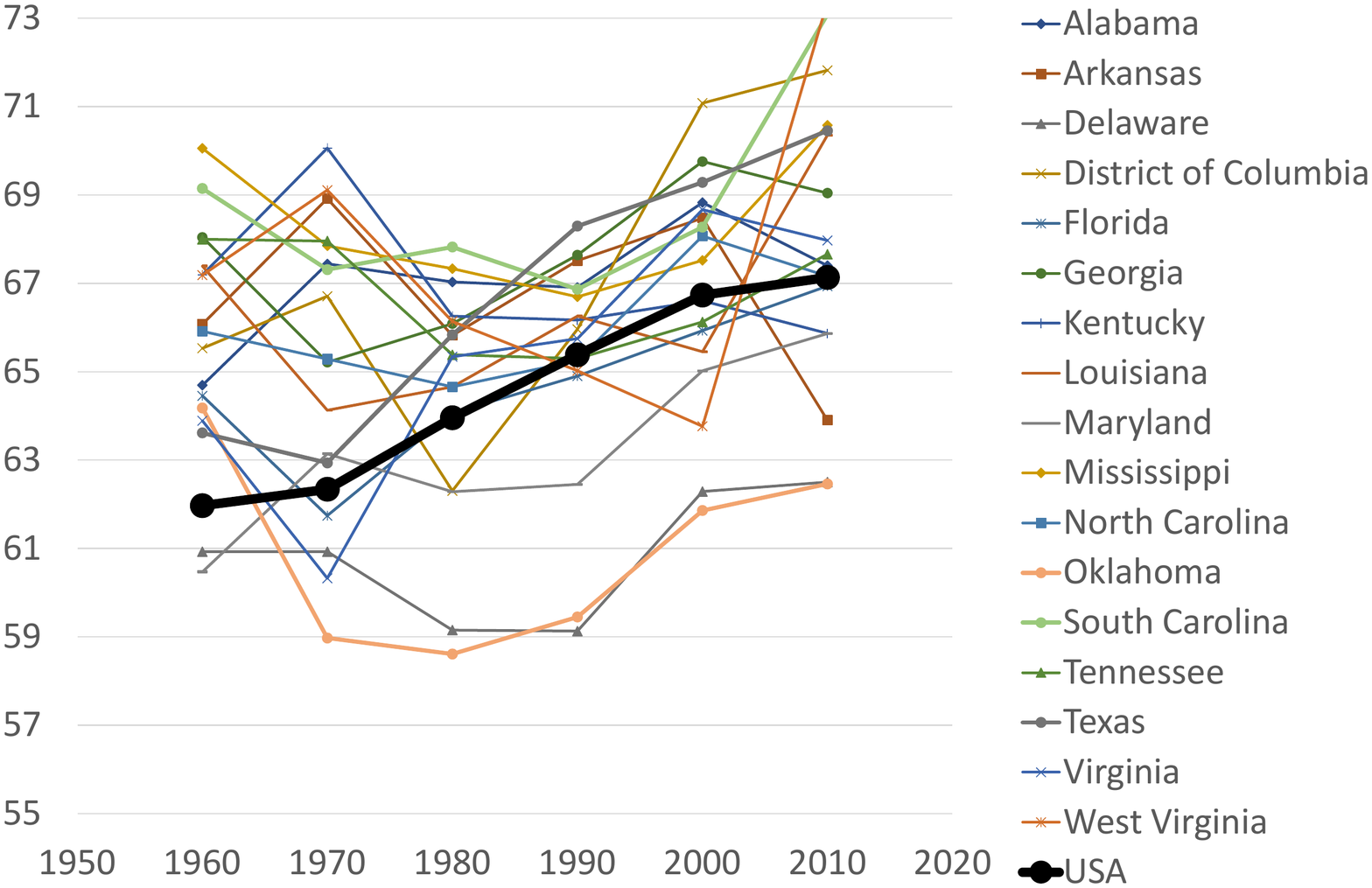}
		\caption{South}
		\label{fig:IPF_S}
	\end{subfigure}%
	\begin{subfigure}{0.7\textwidth}
		\centering
		\includegraphics[width=.9\linewidth]{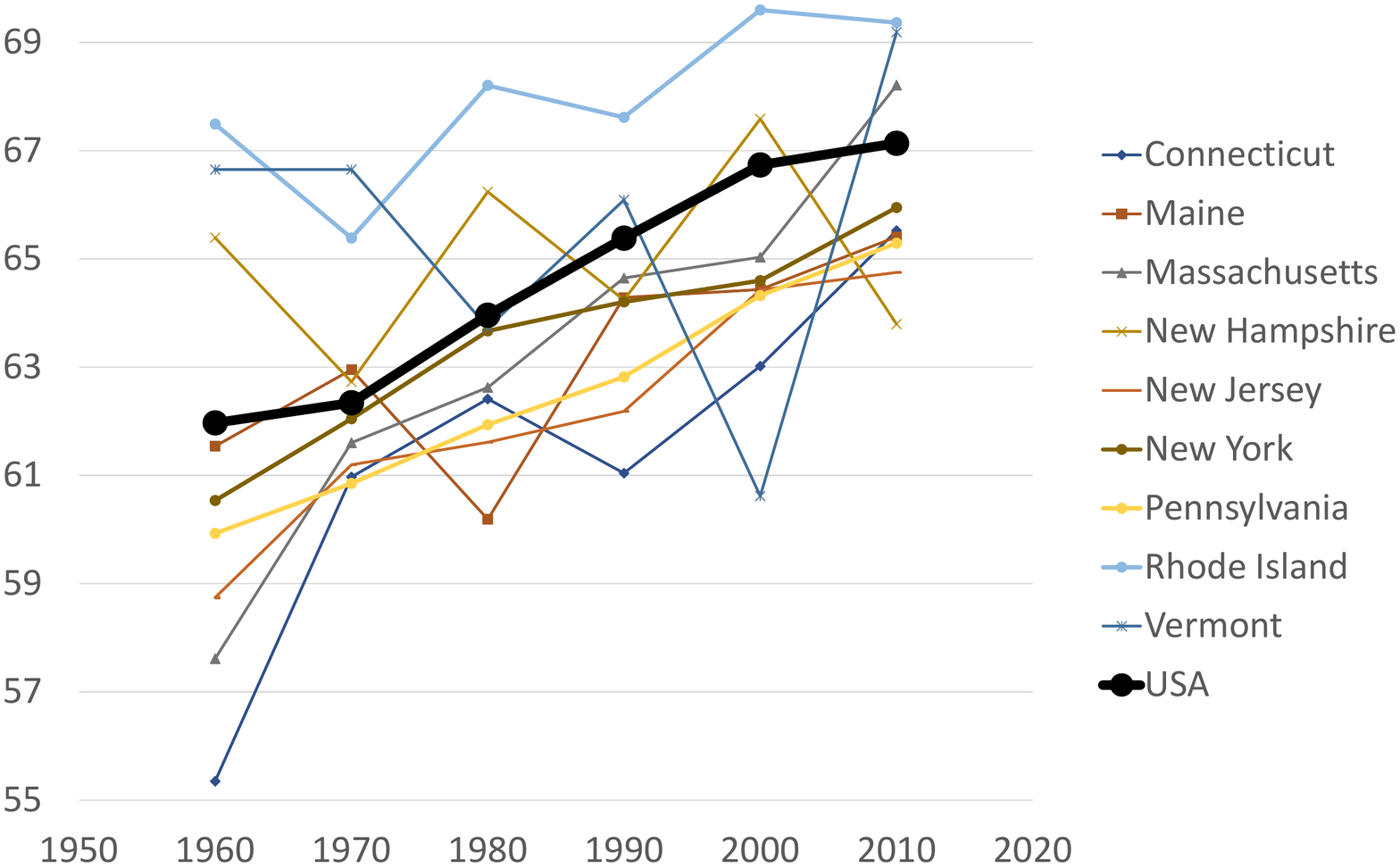}
		\caption{Northeast}
		\label{fig:IPF_NE}
	\end{subfigure}
\end{center}	
	\textit{Source}: author's calculation using decennial census data from IPUMS.\\
\textit{Note}: 	Three education levels are distinguished (no high school degree, high school degree, tertiary education diploma).  
Data on young couples is used, where the age of husbands/male partners is between 30 and 34. 
The values assigned to the year 1960 are the observed, state-specific proportions, or country-level proportion, of educationally homogamous young couples. The country level proportion is calculated as the ratio of the number of homogamous young couples with known residence and the total number of young couples with known residence.   
The values assigned to the years 1970, 1980, 1990, 2000, and 2010  are the corresponding values for 1960 adjusted with the cumulative ceteris paribus effects of changes in homophily across consecutive generations. The ceteris paribus effects are calculated with counterfactual decompositions, where the counterfactual joint educational distribution of couples is constructed with the method named in the caption of this figure. 
	
\end{figure}

\end{landscape}																																			
\addtocounter{figure}{-1}

\newpage
\begin{landscape}																																			
\begin{figure}
	\caption{Trends in educational homophily  obtained with the MEDA algorithm}
	\begin{center}
		
	\begin{subfigure}{0.6\textwidth}
		\centering
		\includegraphics[width=.9\linewidth]{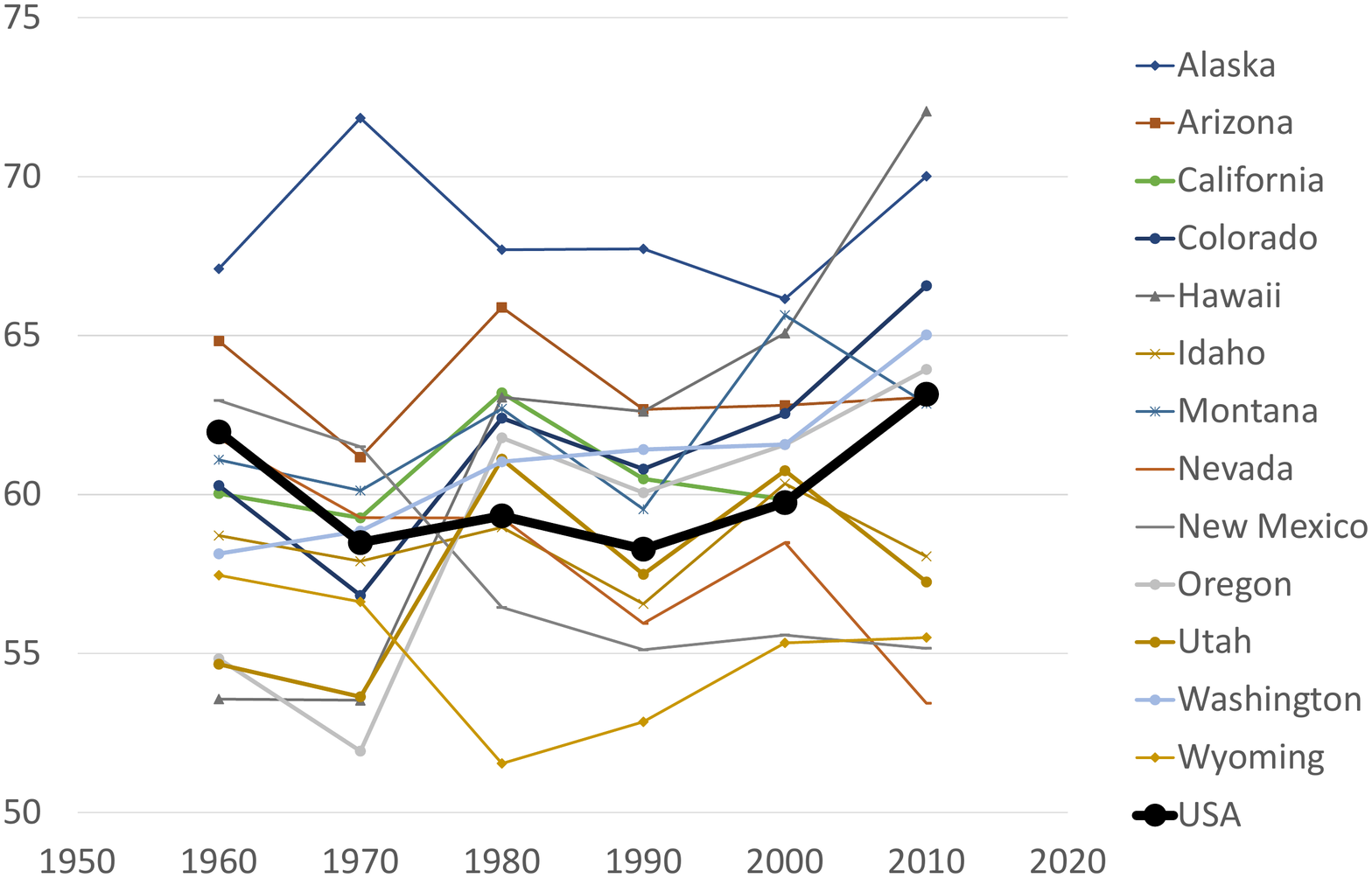}
		\caption{West}
		\label{fig:MEDA_W}
	\end{subfigure}%
	\begin{subfigure}{0.6\textwidth}
		\centering
		\includegraphics[width=.9\linewidth]{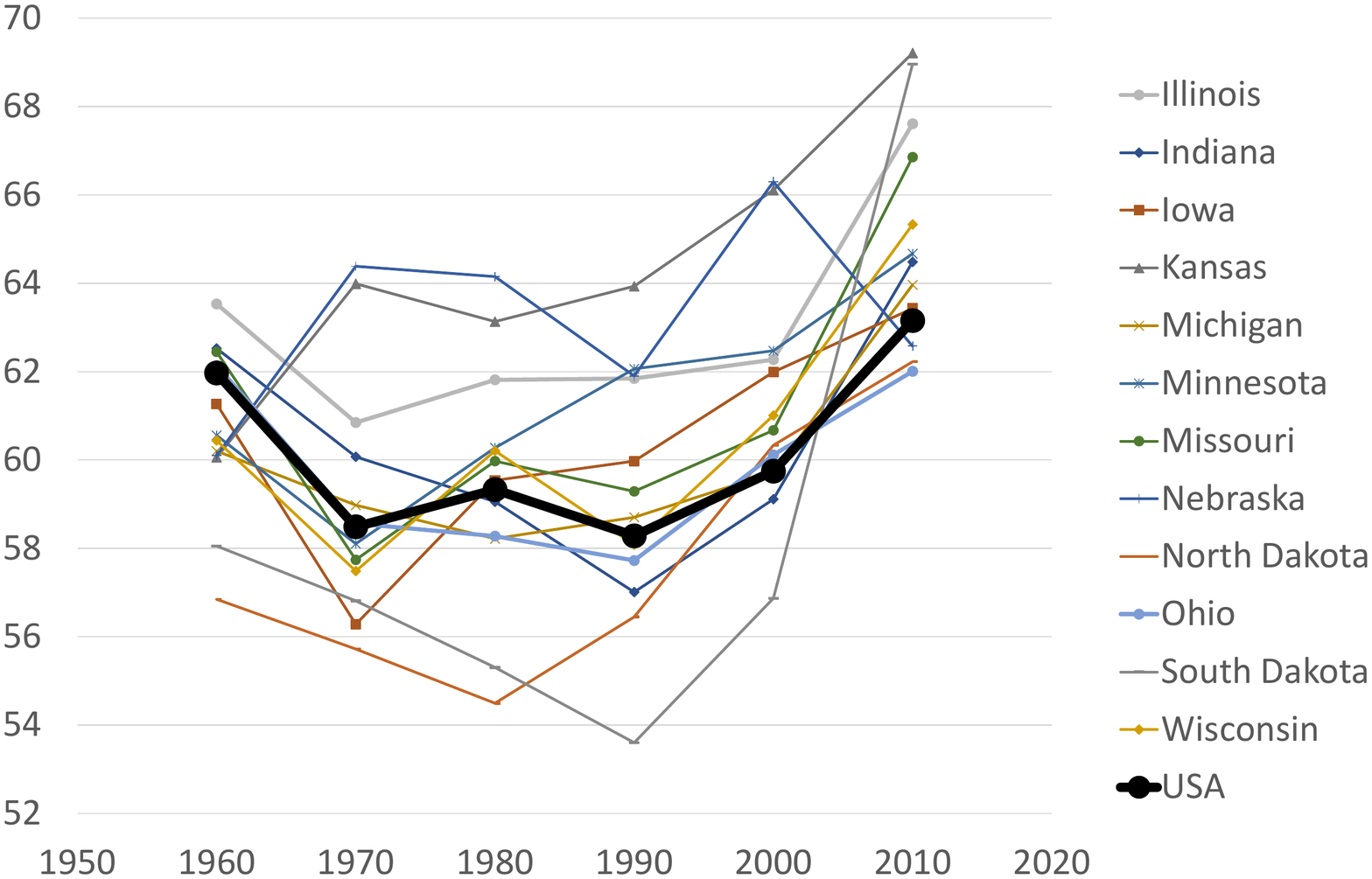}
		\caption{Midwest}
		\label{fig:MEDA_MW}
	\end{subfigure}\\
	
	\begin{subfigure}{0.6\textwidth}
		\centering
		\includegraphics[width=.9\linewidth]{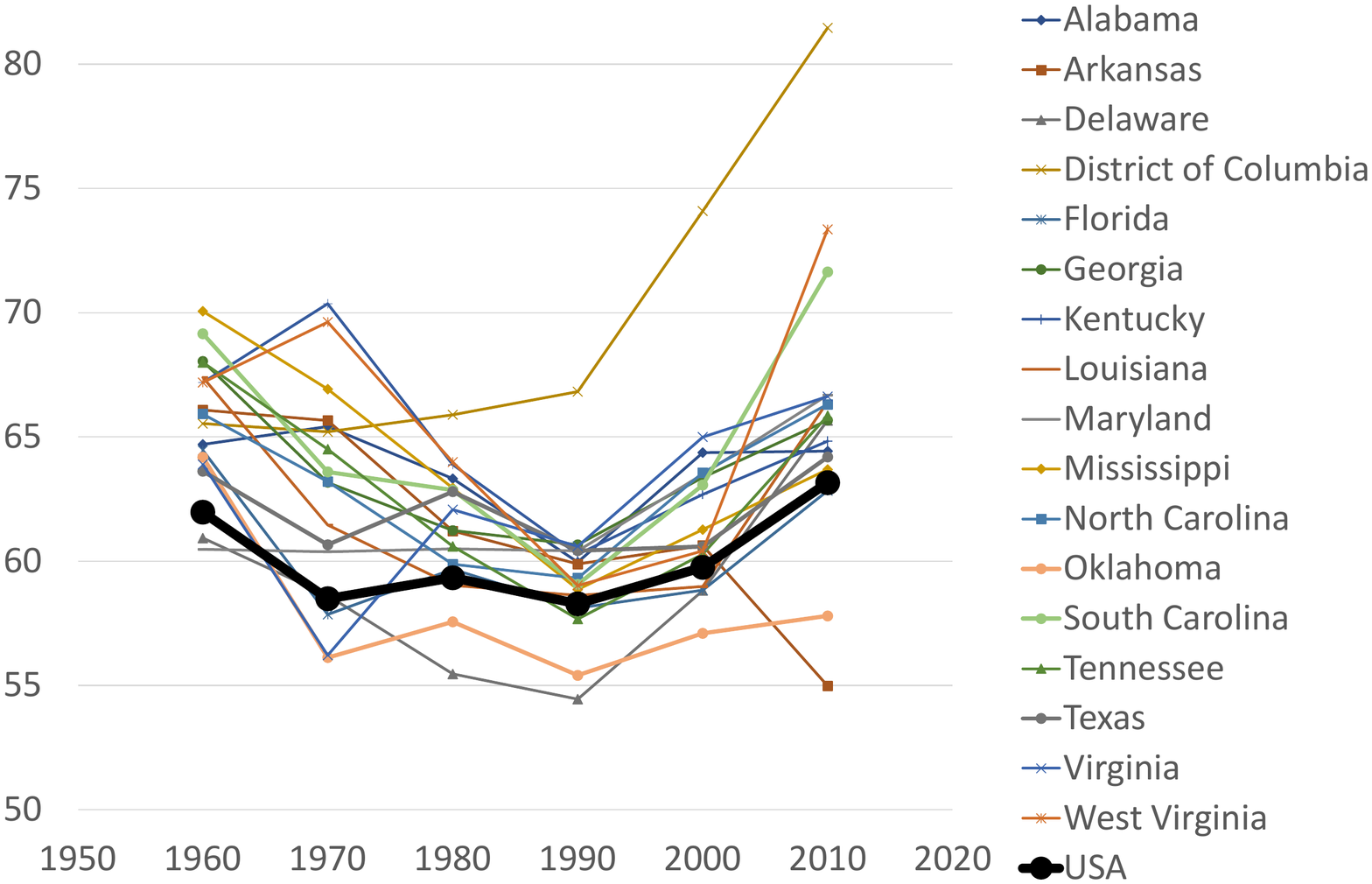}
		\caption{South}
		\label{fig:MEDA_S}
	\end{subfigure}%
	\begin{subfigure}{0.6\textwidth}
		\centering
		\includegraphics[width=1\linewidth]{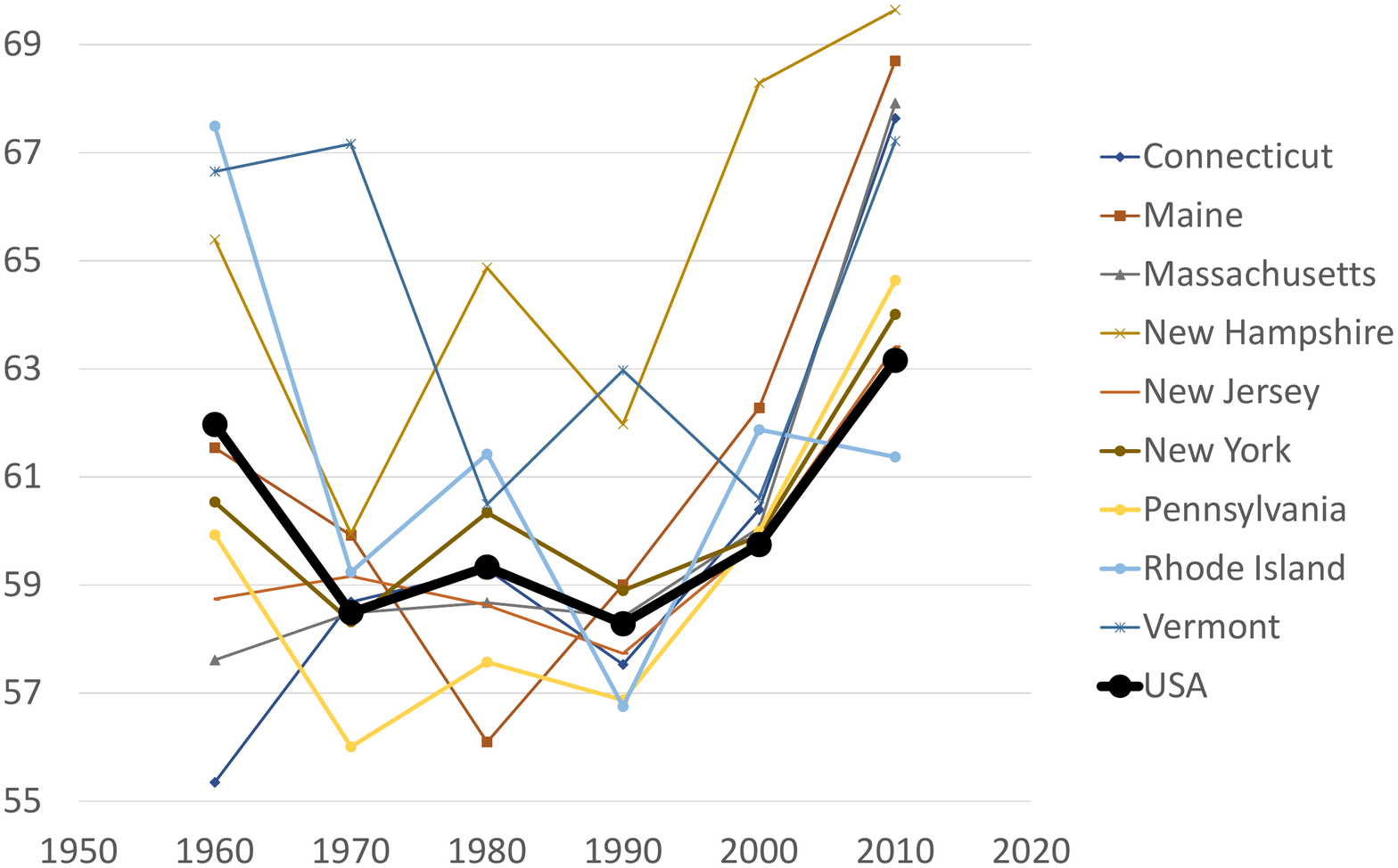}
		\caption{Northeast}
		\label{fig:MEDA_NE}
	\end{subfigure}

	\end{center}	
	\label{fig:MEDA}
	\textit{Note}: same as below Figure \ref{fig:IPF}. 
	
\end{figure}

\end{landscape}																																			
\newpage

\begin{landscape}																																			

\begin{figure}
	\caption{Trends in educational homophily  obtained with the NM-method}
\begin{center}
		
	\begin{subfigure}{.6\textwidth}
		\centering
		\includegraphics[width=.9\linewidth]{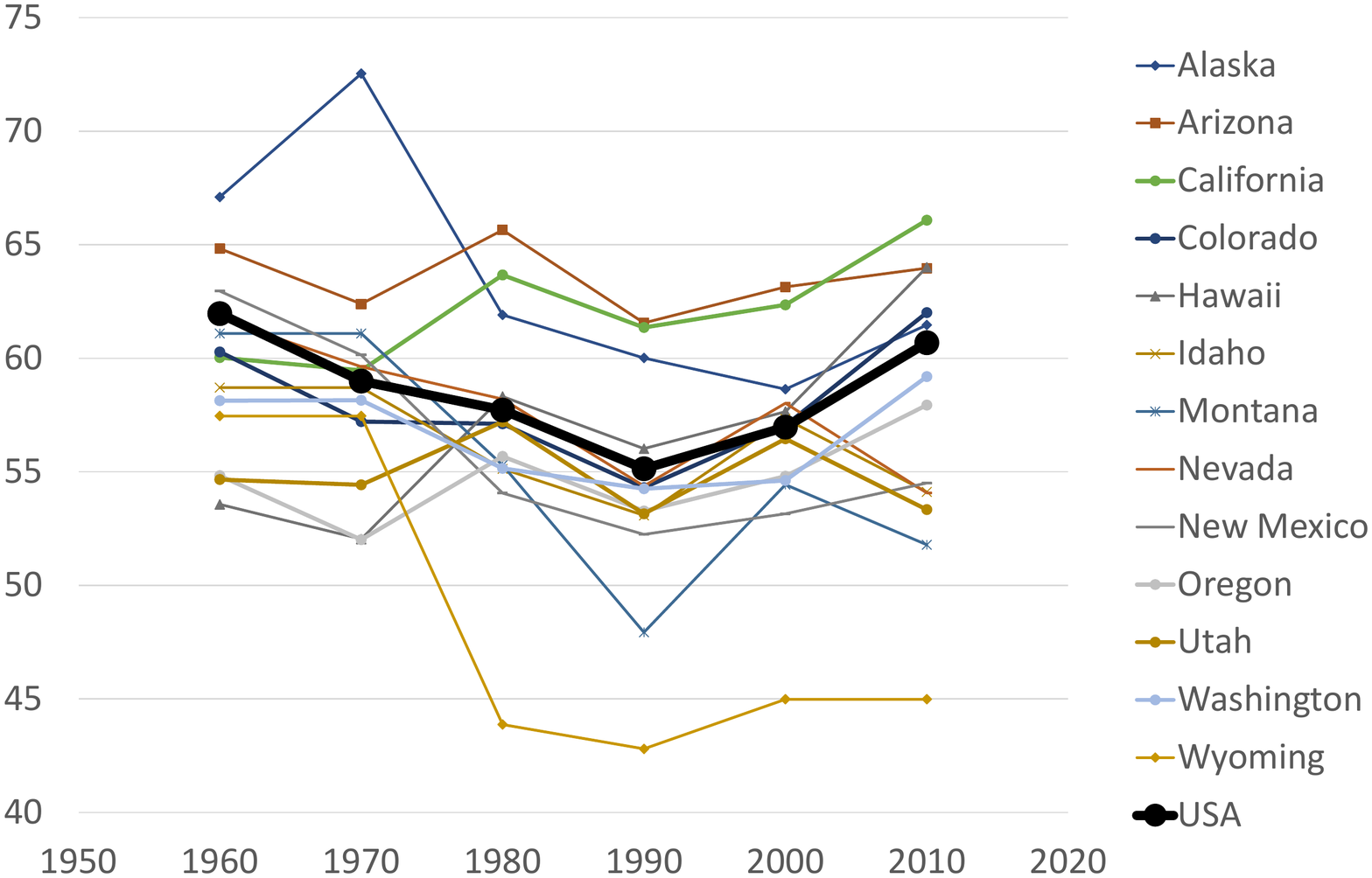}
		\caption{West}
		\label{fig:NM_W}
	\end{subfigure}%
	\begin{subfigure}{.6\textwidth}
		\centering
		\includegraphics[width=.9\linewidth]{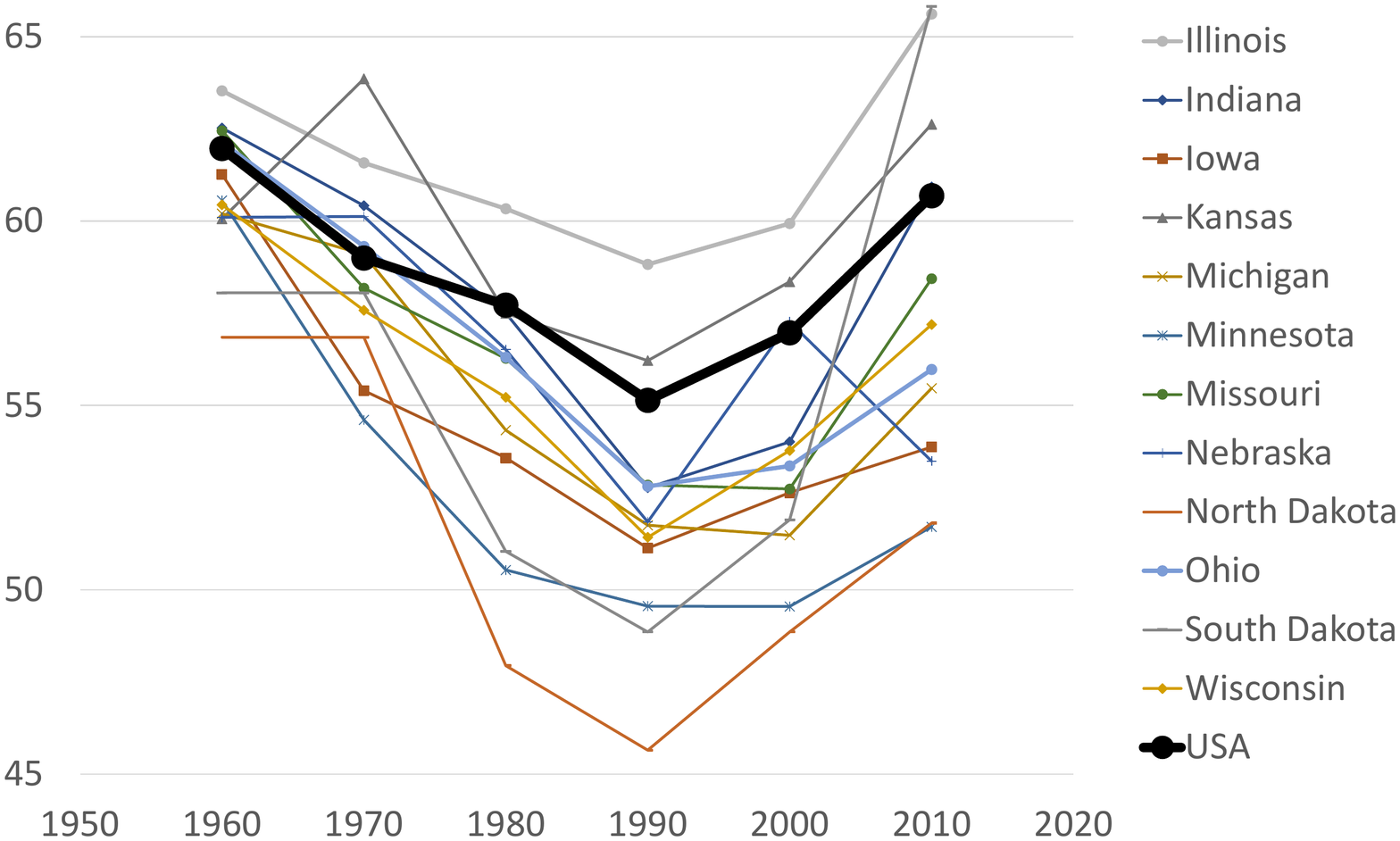}
		\caption{Midwest}
		\label{fig:NM_MW}
	\end{subfigure}\\
	
	\begin{subfigure}{.6\textwidth}
		\centering
		\includegraphics[width=.9\linewidth]{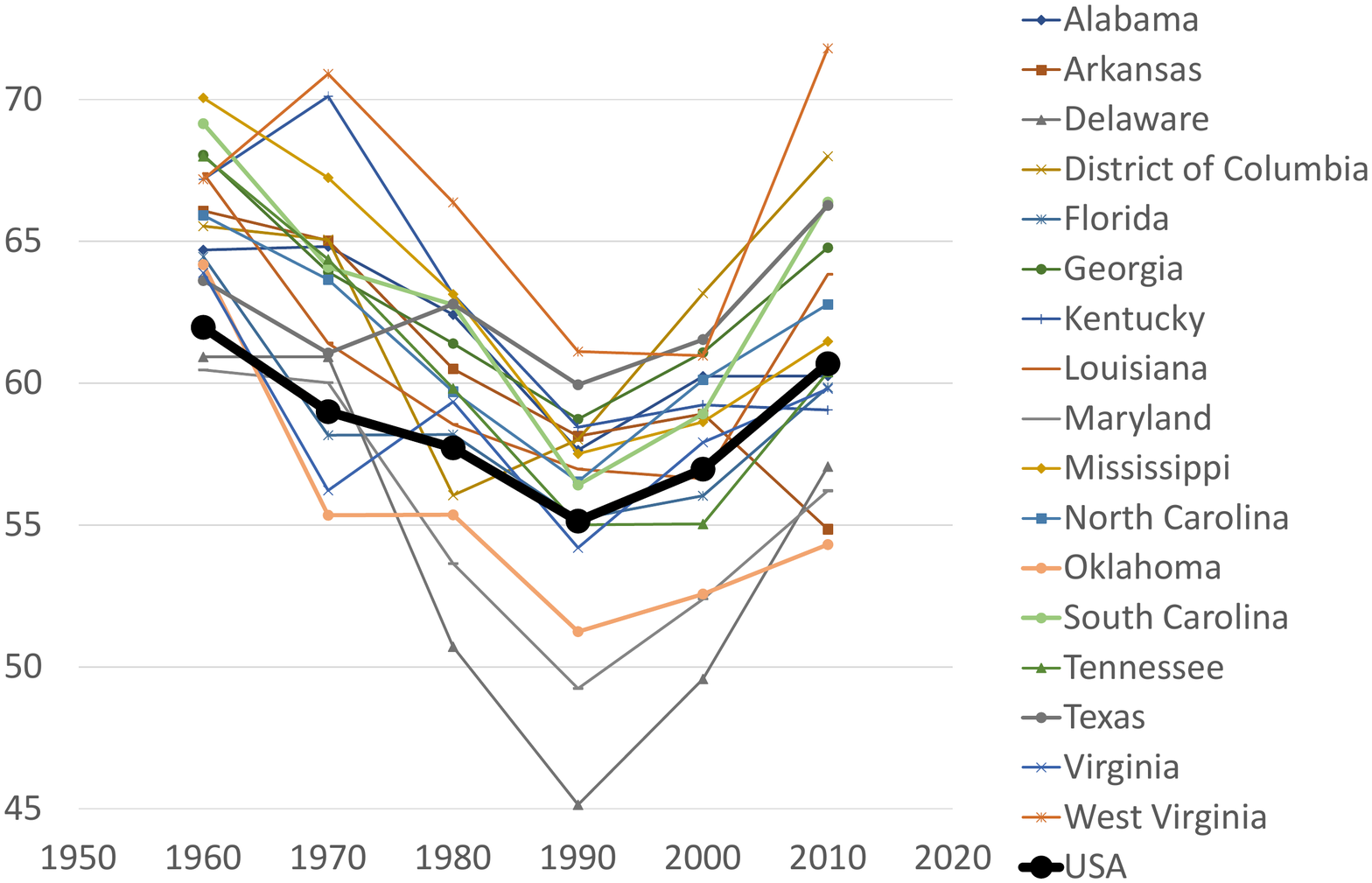}
		\caption{South}
		\label{fig:NM_S}
	\end{subfigure}%
	\begin{subfigure}{.6\textwidth}
		\centering
		\includegraphics[width=1\linewidth]{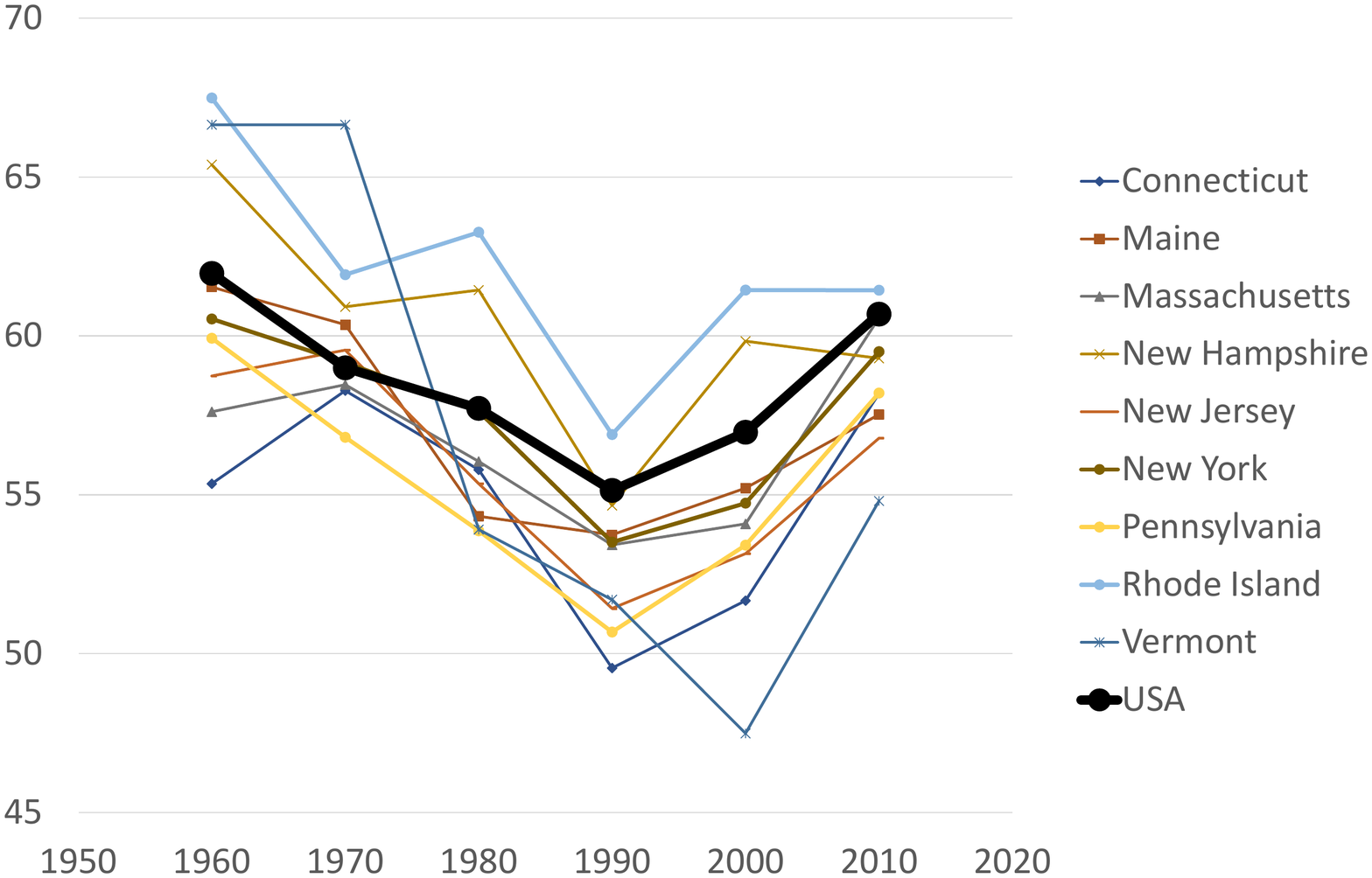}
		\caption{Northeast}
		\label{fig:NM_NE}
	\end{subfigure}

\end{center}	
	\label{fig:NM}
		\textit{Note}: same as below Figure \ref{fig:IPF}. 
	
\end{figure}

\end{landscape}																																			

\newpage


	\begin{figure}
		\caption{Trends in educational homophily in the US obtained with either  the CS model, or the IPF, or the NM, or the GNM}
\begin{center}

		\begin{subfigure}{.8\textwidth}
	\centering
	\includegraphics[width=.9\linewidth]{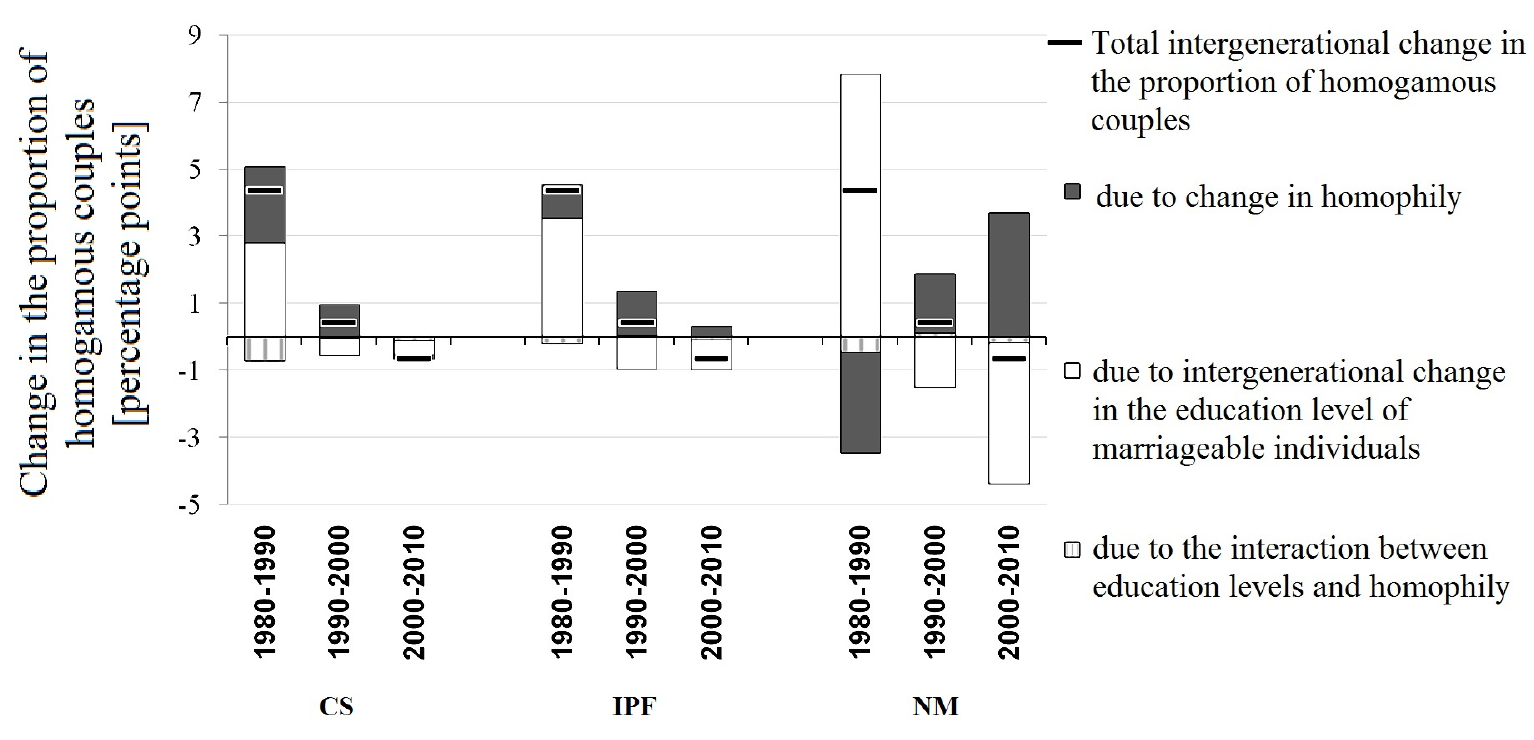}
	\caption{CS, IPF and NM}
	\label{fig:CSIPFNM}
\end{subfigure}\\
			
		\begin{subfigure}{.8\textwidth}
			\centering
			\includegraphics[width=.9\linewidth]{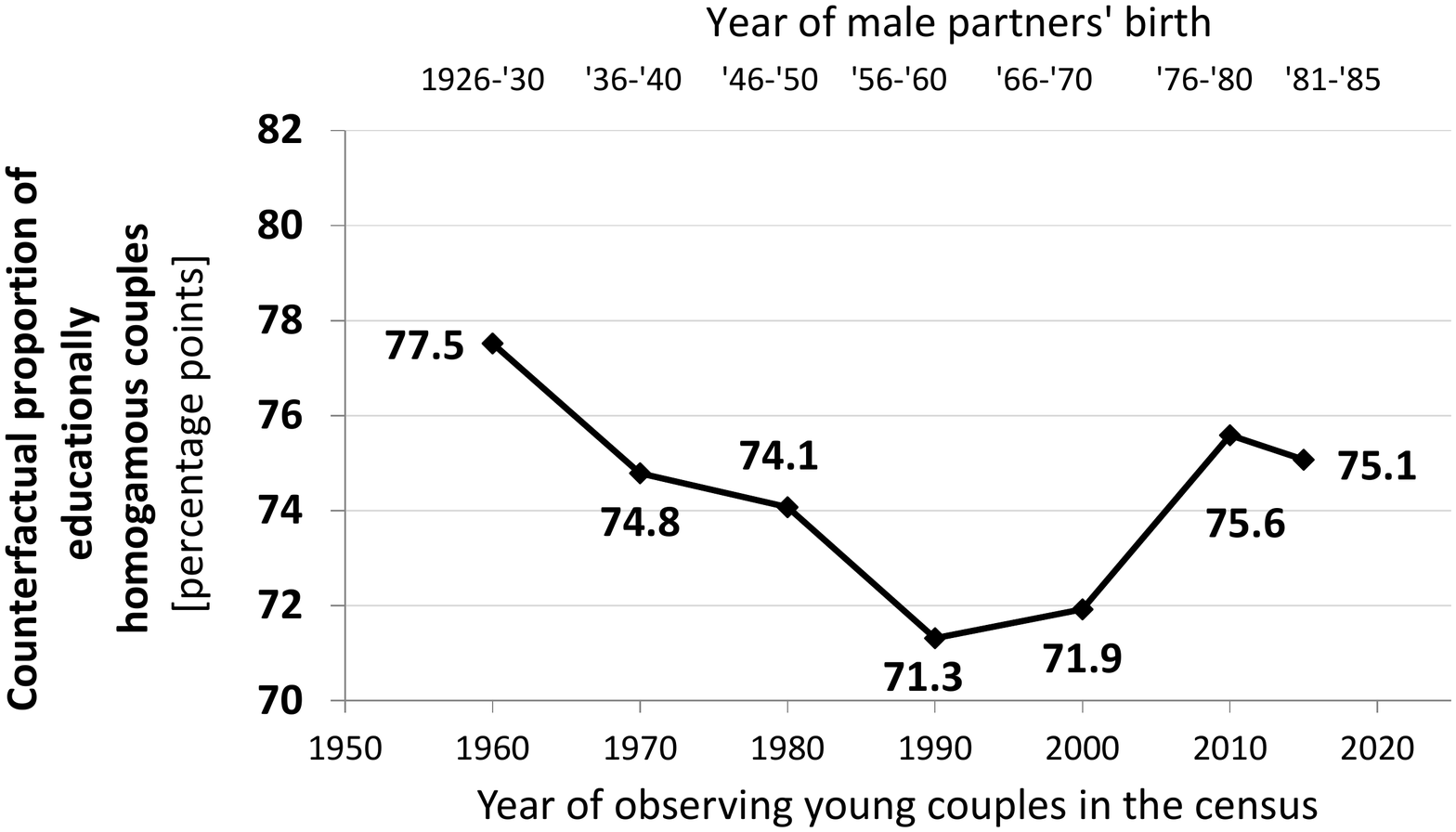}
			\caption{GNM}
			\label{fig:GNM}
		\end{subfigure}

\end{center}		
\label{fig:rest}
		\textit{Note}: partial reproduction of Figure 1b in \cite{NaszodiMendonca2021} and Figure A6 in \cite{NaszodiMendonca2023_RACEDU}. 
		The black bars of Subfigure \ref{fig:CSIPFNM} show that the CS, similar to the IPF,  attributes an increase in the prevalence of homogamy to the changing homophily of American late Boomers (observed in 1990) relative to the  early Boomers (observed in 1980) and it finds no change in homophily across the early and late GenXers (observed in 2000 and 2010, respectively). By contrast, the NM and the GNM attribute a decrease in the prevalence of homogamy to the changing homophily from the early Boomers to the late Boomers and an increase in the prevalence of homogamy to the changing homophily from the early GenXers to the late GenXers.   
				
				\label{fig:CSGNM}
				
	\end{figure}

\end{appendices}

\end{document}